\documentclass[10pt,conference]{IEEEtran}

\AtBeginEnvironment{tabular}{\fontsize{32}{36}\selectfont}
\IEEEoverridecommandlockouts

\AtBeginDocument{%
  \providecommand\BibTeX{{%
    \normalfont B\kern-0.5em{\scshape i\kern-0.25em b}\kern-0.8em\TeX}}}

\newif\ifdraft
\drafttrue %shows boldifications

\usepackage[numbers,sort&compress,square]{natbib}
\usepackage{soul}
\usepackage{graphicx}
\usepackage{xurl}
\usepackage{booktabs}
\usepackage{multibib}
\usepackage{natbib}
\usepackage{hhline}
\usepackage{xcolor}
\usepackage{graphicx}
\usepackage{colortbl}
\usepackage{multirow}
\usepackage{tabularx}
\usepackage{amsmath}
\usepackage{makecell}
\usepackage{textgreek}
\usepackage{tabularx}
\usepackage{siunitx}
\usepackage{stfloats, caption}%
\usepackage{tablefootnote}

\newcolumntype{Y}{>{\centering\arraybackslash}X}
\usepackage{bookmark}
\usepackage{textcomp}
\usepackage{xcolor}
\usepackage{microtype}
\usepackage{booktabs}
\usepackage{etoolbox}
\usepackage{enumitem} 
\usepackage{url}
\usepackage{ragged2e}

\usepackage{amsmath} % For text command

\newcommand{\rqone}{How are interpersonal challenges associated with the feeling of welcomeness in OSS?}
\newcommand{\rqtwo}{How does the association between interpersonal challenges and feeling welcome differ for those who are in gender, race, or disability minorities?}

\newcommand{\rqthree}{How differently do those in gender, race, or disability minority face interpersonal challenges?}

\author{\IEEEauthorblockN{Anonymous Authors}}

\newcolumntype{P}[1]{>{\raggedright\let\newline\\\arraybackslash\hspace{0pt}}m{#1}}

\newcolumntype{C}[1]{>{\centering\let\newline\\\arraybackslash\hspace{0pt}}m{#1}}

\newcommand\hypothesis[2]{\vspace{0.5em} \noindent \hangindent=1em \textbf{Hypothesis #1 (H#1).} {#2}\vspace{0.5em}}

\newcommand{\MyBox}[1]{\vspace{3mm}\noindent\framebox[\columnwidth][c]{\parbox[b]{0.95\columnwidth}{ #1 }}\vspace{3mm}}

\begin{document}

%% allowing the author to define a "short title" in page headers.
%% KS Removed newline because pushes title into three lines which is suboptimal.
%\title{Interpersonal Challenges and Not Feeling Welcome in Open Source Software}
%\title{Unlocking the Trifecta: Examining Interpersonal Challenges, Minority Representation, and Belonging in OSS Communities}
%\title{Untangling the Trifecta of OSS misery: Interpersonal Challenges, Minority Representation, and Sense of Belonging}
%\title{The Domino effect: Investigating Interpersonal Challenges, Minority Representation, and Belonging in OSS Communities}
%\title{The Domino Effect on attrition in OSS: Investigating Interpersonal Challenges, Minority Representation, and Being Welcomed}
%\title{Unpacking Retention in OSS: Effects of Interpersonal Challenges, Minority Representation, and Feeling Welcome}
%\title{Unpacking the Effects of Challenges, Minority Representation, and Feeling Welcome in OSS}
%\title{Untangling the Skeins of Challenges Faced, Minority Representation, and Feeling Welcome in OSS}
%\title{Unpacking Which Challenges Matter and To Who in Feeling Welcome in OSS}
%\title{Investigating the Differential Impact of Challenges on Minority Groups and Feeling Welcome in OSS}
\title{Investigating the Impact of Interpersonal Challenges on Feeling Welcome in OSS}

\author{
\IEEEauthorblockN{
Bianca Trinkenreich\IEEEauthorrefmark{1}, Zixuan Feng\IEEEauthorrefmark{2}, Rudrajit Choudhuri\IEEEauthorrefmark{3}, Marco Gerosa\IEEEauthorrefmark{3}, Anita Sarma\IEEEauthorrefmark{2}, Igor Steinmacher\IEEEauthorrefmark{3}}
\\
\vspace{-4mm}
\IEEEauthorblockA{\IEEEauthorrefmark{1}Colorado State University, United States, \{bianca.trinkenreich\}@colostate.edu}
\IEEEauthorblockA{\IEEEauthorrefmark{2}Oregon State University, United States, \{fengzi,choudhru, anita\}@oregonstate.edu}
\IEEEauthorblockA{\IEEEauthorrefmark{3}Northern Arizona University, United States, \{marco.gerosa,igor.steinmacher \}@nau.edu}
 \vspace{-5mm}
}

\maketitle

\begin{abstract}
%\textbf{Context:} 
The sustainability of open source software (OSS) projects hinges on contributor retention. Interpersonal challenges can inhibit a feeling of welcomeness among contributors, particularly from underrepresented groups, which impacts their decision to continue with the project. How much this impact is, varies among individuals, underlining the importance of a thorough understanding of their effects. Here, we investigate the effects of interpersonal challenges on the sense of welcomeness among diverse populations within OSS, through the diversity lenses of gender, race, and (dis)ability. We analyzed the large-scale Linux Foundation Diversity and Inclusion survey (n = 706) to model a theoretical framework linking interpersonal challenges with the sense of welcomeness through Structural Equation Models Partial Least Squares (PLS-SEM). We then examine the model to identify the impact of these challenges on different demographics through Multi-Group Analysis (MGA). Finally, we conducted a regression analysis to investigate how differently people from different demographics experience different types of interpersonal challenges. Our findings confirm the negative association between interpersonal challenges and the feeling of welcomeness in OSS, with this relationship being more pronounced among gender minorities and people with disabilities. We found that different challenges have unique impacts on how people feel welcomed, with variations across gender, race, and disability groups. We also provide evidence that people from gender minorities and with disabilities are more likely to experience interpersonal challenges than their counterparts, especially when we analyze stalking, sexual harassment, and doxxing.
Our insights benefit OSS communities, informing potential strategies to improve the landscape of interpersonal relationships, ultimately fostering more inclusive and welcoming communities.

\end{abstract}


\vspace{-5px}
\section{Introduction}
\label{sec:introduction}
%\boldification{OSS is voluntary and need people to feel part of the community to be there. }

A core principle of Open Source Software (OSS) is its openness---anyone from anywhere in the world can contribute, irrespective of their background. However, despite this apparent openness, OSS projects have a severe lack of diversity. For example, data reveals that less than 10\% of the contributors are women~\cite{trinkenreich2022women}, and rates are also low for non-whites (16.6\%)~\cite{nadri2021relationship} and people who have disabilities (17\%)~\cite{lfdiversitysurvey}. This lack of diversity is not just a social justice problem, it also undermines the innovation and sustainability of OSS projects~\cite{vasilescu2015gender, bosu2019diversity}.
%The inclusion of a diverse set of contributors can significantly benefit OSS projects \cite{vasilescu2015gender}, and contributing to OSS can advance contributors' careers \cite{gerosa2021shifting,trinkenreich2021pot,roberts2006understanding}. Therefore, understanding the challenges faced by minorities and devising strategies to overcome them is crucial.

%The challenges contributors face exacerbate the already difficult task of joining OSS and can drain contributors' enthusiasm and motivation~\cite{steinmacher2014attracting}. This difficulty is only one of the problems hindering diversity in OSS \cite{vasilescu2015perceptions,lee2019floss}. 
%The Linux Foundation Research ran a survey on 2021 that was answered by 2,350 people about the state of diversity and Inclusion in OSS \cite{dei_survey}. Although the report indicated that 82\% of respondents feel welcome in OSS, it also highlights that the sentiments differ when segmenting the results by demographics. Those who do not feel welcome are primarily from underrepresented groups, including gender minorities, individuals with disabilities, and racial background minorities .

Literature has also extensively explored the multiple facets of challenges faced by OSS contributors \cite{steinmacher2015social,Steinmacher.Conte.ea_2015HICSS,jensen2011joining, hannebauer2017relationship}, including barriers faced by newcomers~\cite{steinmacher2015social}, one-time contributors \cite{pinto2016more}, and mentors \cite{balali2018mentor}. %, and contributors from diverse populations~\cite{trinkenreich2022women, nadri2021insights, rodriguez2021perceived}.
Prior work has investigated how a Sense of Virtual Community (SVC) in OSS is created by intrinsic factors such as membership, identity, and attachment to a group and affects their sense of belonging \cite{trinkenreich2023belong}. However, \cite{trinkenreich2023belong} does not tie this feeling of SVC to the challenges that individuals face, which are the antecedents of SVC.
More specifically, studies have identified interpersonal challenges as a major obstacle to contributing to OSS \cite{storey2016social}, especially for those who are a minority in OSS~\cite{trinkenreich2022women, rodriguez2021perceived, nadri2021insights}.

%What is missing, however, is an understanding of the extent to which these interpersonal challenges hinder the creation of a feeling of welcomeness in the project, which in turn impacts contributors' decision to continue.

What is missing, however, is an understanding of how these interpersonal challenges hinder the contributors' feeling of welcomeness in the project, which in turn impacts their decision to continue contributing. This oversight is not trivial; the feeling of welcomeness is key to sustained engagement, particularly in environments built on open collaboration principles~\cite{forte2013defining}. This brings us to our first research question: 
\textbf{RQ1: }\rqone

%\\
% \vspace{-5px}
% \noindent\textbf{RQ1: }\rqone\\
% \vspace{-5px}

Research has also shown that individuals in the minority in OSS often face a higher frequency of interpersonal challenges such as toxic communication or gender stereotyping, which in turn impacts their turnover \cite{lee2019floss,balali2018mentor,trinkenreich2022women}. However, the extent to which interpersonal challenges affect the feeling of welcomeness among minority communities is yet to be determined. Additionally, our worldview is an amalgamation of our individual backgrounds and lived experiences. Consequently, various interpersonal challenges may exert disparate influences on individuals belonging to distinct demographic groups. 
Thus far, no research has quantified the differential effects of various challenges on feeling welcome nor on the interpersonal challenges across demographic dimensions. This leads us to our other two research questions: 
\textbf{RQ2: }\rqtwo 
\textbf{ RQ3: }\rqthree

%\\

% \vspace{-5px}
% \noindent \textbf{RQ2: }\rqtwo\\
% \noindent \textbf{RQ3: }\rqthree\\
% \vspace{-5px}

The current absence of a comprehensive understanding that delineates the impact of challenges on the feeling of welcomeness across various population segments implies that endeavors by OSS communities to retain a diverse pool of contributors are likely to fail. 

Our study closes this gap through a large-scale (n=706), mixed-methods analysis of Linux Foundation survey data ~\cite{dei_survey}. 
Specifically, we build a theoretical model using Partial-Least Squares Structural Equation Modeling (PLS-SEM) to determine and delineate the different types of interpersonal challenges, their extent, and their impact on feeling welcome (RQ1). We then model the differential impact of the various interpersonal challenges on demographic dimensions of gender, race, and disabilities through Multi-Group Analysis (RQ2). Following, we modeled individual logistic regressions to evaluate the differences in experiencing interpersonal challenges across demographic dimensions of gender, race, and disabilities (RQ3).

Our study follows a concurrent embedded strategy \cite{creswell2017research} and uses qualitative analysis of the open-ended survey responses about reasons to feel (or not feel) welcome to understand the participants' specific experiences that contributed to their feeling (or lack) of welcomeness in the community. This serves as a form of data triangulation to provide rigor to our findings. 

Our theoretical model empirically shows that experiencing interpersonal challenges is negatively associated with feeling welcome in OSS. Moreover, it highlights the nuanced nature of how different challenges impact different demographics differently. For example, in our data, while race minorities felt less welcome, interpersonal challenges had a strong association with not feeling welcome for gender minorities and people with disabilities, but not for race. %Respondents from race minorities felt less welcome when experiencing unwelcoming language, but this challenge did not significantly impact feeling welcomed for gender minorities or people with disabilities in our data. 

%Doxxing impacts feeling welcome only for gender minorities, not for race minorities nor people with disabilities. 
%Both minorities and majorities feel discriminated against, and 

%Stereotyping had no significant effect on the feeling of welcomeness for gender, race, or people with disabilities. However, challenges such as stalking, sexual harassment, and lack of response or rejection of contributions made the three investigated minority groups feel less welcome than their counterparts.

By understanding how specific challenges impact diverse contributors' feelings of welcomeness, our study not only advances the academic literature on the field of behavioral software engineering but also holds practical implications. The results are paramount to inform OSS communities on the specific aspects to focus on to foster diversity in OSS, which in turn can result in innovation, and drive the sustainability of OSS projects.
Our contributions offer valuable insights that may go beyond OSS, informing broader software industry efforts on creating inclusive and equitable spaces.

\vspace{-3pt}
\section{Related Work}
\label{sec:related}
\vspace{-2pt}

%In this section, we review the existing literature on diversity and inclusion in OSS, interpersonal challenges in OSS, and the sense of belonging within OSS, discussing how our research complements these studies.

%Such attention and efforts inspire hope in OSS communities; according to a CNCF survey, seven out of ten respondents agree OSS is becoming more inclusive \cite{cncf2021diversity}. However, the structures and culture of the entire OSS communities are much complicated, and reality can be cruel. 

\noindent\textbf{Diversity and Inclusion in OSS.} Diversity and inclusion in OSS have received attention in recent years \cite{feng2023state, feng2023state1, trinkenreich2022women, trinkenreich2020hidden, trinkenreich2021pot, trinkenreich2023belong, vasilescu2015gender,lee2019floss, bosu2019diversity}. Most of the literature related to minorities and underrepresentation in OSS is focused on women~\cite{vasilescu2015gender,trinkenreich2022women,trinkenreich2020hidden, bosu2019diversity, lee2019floss,Mendez2018GE}. A literature survey by \citet{trinkenreich2022women} identified that women are poorly represented and aggregates eight significant challenges women face when contributing to OSS. 
%Bosu and Sultana amplify this voice by showing that despite the growing awareness and efforts to improve diversity and inclusion in OSS projects, significant challenges remain, as women continue to face barriers to participation. 
By analyzing repositories, ~\citet{vasilescu2015gender} showed that projects with higher gender diversity within their development teams tend to experience more effective collaboration and productivity. Similarly, a survey by ~\citet{lee2019floss} highlighted notable perceptions of gender bias and inclusivity issues within the OSS community. Participants acknowledged barriers to women's participation and emphasized the need for better practices to foster inclusivity.

However, fewer studies have focused on other underrepresented populations in OSS besides gender. Previous research claimed for more studies about racial diversity in software engineering \cite{rodriguez2021perceived}. Race minority was evidenced as underrepresented in OSS \cite{guizani2022perceptions, huang2021leaving}. In this context, \citet{nadri2021insights} found that perceptibly Black and Asian/Pacific developers faced higher contribution rejection rates than those from perceptibly White developers. A literature review by \citet{rodriguez2021perceived} indicates limited knowledge about the impact of race, nationality, and disability in OSS. Despite this, respondents in a recent survey reported that their disability influences their emotions and perceptions of others' feelings \cite{apache2023dei}.

\noindent\textbf{Interpersonal challenges in OSS.} ``Open source has a reputation for being aggressive"~\cite{DiversityOSS2023}. Toxicity and incivility are prevalent issues in OSS discussions and code reviews, adversely affecting contributor engagement, collaboration, and overall community health ~\citet{miller2022did, ferreira2021shut}. Contributors who are underrepresented in OSS are experiencing even worse collaboration environments. Harassing verbal comments, sexual jokes, insults, and bullying are not uncommon \cite{powell2010gender, vasilescu2015perceptions}. \citet{singh2022discrimination} found that less than 5\% of online communities are ``safe'' to women contributors, free of sexism and discrimination. However, from a quantitative analysis of 355 OSS software package websites, only 10\% of them have a Code of Conduct or relevant rules \cite{singh2021codes}. Such prevalent interpersonal challenges are creating barriers that make contributors feel unwelcome in OSS, especially for contributors who are underrepresented.

% Raman et al.~\cite{raman2020stress} evidenced that contributors to open source software projects frequently experience stress and burnout, and emphasized the need for strategies to identify, understand, and mitigate these issues to support the well-being of contributors. 
\noindent\textbf{Welcomeness in OSS. } 
%‘Feeling welcome’ attends to the sense of being invited to use or participate on a given group being free of judgment.
Feeling welcome antecedes belonging \cite{hailey2022racialized}. \citet{gill2018suppression} describes welcomeness as more than a mere acknowledgment of arrival or permission for entry. It embodies a deeper sense of emotional and relational connection between the welcomer and the newcomer, implying a willingness to engage and suggesting a deeper level of engagement beyond mere permission. It involves an exchange of emotions that transcend bureaucratic processes~\cite{gill2018suppression}.

We introduce the concept of welcomeness in OSS, which focuses on conveying a genuine positive reception of newcomers as contributors to the project. While onboarding primarily addresses the procedural and technical requirements necessary to contribute, welcomeness emphasizes the interpersonal dynamics that foster an engaging atmosphere characterized by mutual respect and connection.

Welcomeness is a construct investigated in the context of the inclusion of minorities in other areas, like, A~\cite{foor2007wish}, Education~\cite{bopp2017you,pitterson2022helping}, Sports~\cite{bopp2017you}, International Relations~\cite{williams2015beyond}, and Healthcare~\cite{wen2007homeless}. While there is no previous work specifically focusing on welcomeness in software engineering, it antecedes the sense of belonging~\cite{hailey2022racialized}---which was the focus on recent studies in software engineering teams context~\cite{gren2022makes, trinkenreich2023belong, trinkenreich2024unraveling}. 

The need for belonging is crucial in the OSS domain, as shown in studies of OSS contributors~\cite{trinkenreich2021pot,trinkenreich2023belong}. More specifically, close to the present work, ~\citet{trinkenreich2023belong} showed that the sense of virtual community plays a critical role in the contributors' continued engagement and contributions. Trinkenreich's study highlights that factors such as shared experiences, support, and meaningful interaction within the community are vital in fostering a sense of belonging among contributors, thereby influencing their commitment to the OSS environment. Therefore, the sense of belonging emerges as a key factor in retaining personnel, which is vital for the sustainability of OSS projects. 

While previous research extensively investigates the various aspects analyzed in this study, it does not connect aspects of belonging with the interpersonal challenges faced by contributors nor consider the feeling of welcomeness as a distinct construct. Our present work complements the existing literature by evidencing the interplay between interpersonal challenges and welcomeness---which anticipates belonging \cite{hailey2022racialized}---and their effect on gender, race, and disabilities minorities.

% Additionally, non-White developers were more commonly left non-merged without explanation compared to their perceptibly White counterparts.

% Stereotyping in OSS may form a bias against who submitted a code and had been investigated across perceived gender and race. Despite women sometimes having merge acceptance rates similar to or slightly higher than men, Terrell et al. \cite{terrell2017gender} found a bias against women's contributions when gender is disclosed. However, when gender was not identified, women generally had slightly higher pull-request acceptance rates than their counterparts, regardless of experience level. Nadri et al. 

%\textbf{SOB}

\vspace{-5pt}
\section{Research Design}\label{sec:researchdesign}
\vspace{-5pt}

\begin{figure*}[!btp]
\centering
\includegraphics[width=\textwidth]{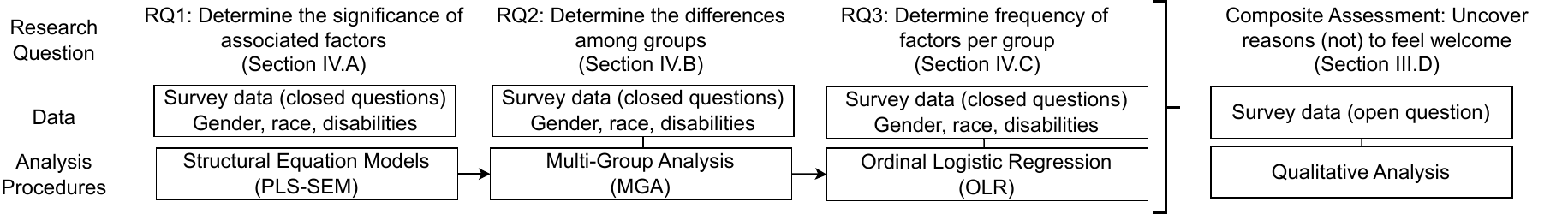}
\vspace{-20px}
\caption{\footnotesize Research design following a Concurrent Embedded Mixed Methods Strategy} \label{fig:research_design}
\vspace{-15px}
\end{figure*}

Fig.~\ref{fig:research_design} presents an overview of the research design. We followed a Concurrent Embedded Mixed Methods Strategy \cite{creswell2017research} with a dominant quantitative component.

Answering \textsc{RQ1} involves using Structural Equation Models - Partial Least Squares to model how interpersonal challenges are associated with the feeling of welcomeness. In RQ2, we investigated the heterogeneity in the model through Multi-Group Analysis (MGA) and established whether the association of the variables identified in RQ1 varied according to gender, race or ethnicity, and abilities. \textsc{RQ3} identifies the frequency of each demographic group experiencing each interpersonal challenge using Ordinal Logistic Regression. Finally, we qualitatively analyzed the open-ended survey responses about why contributors (do not) feel welcome in OSS as a form of data triangulation and conducted a follow-up study to collect the practitioners' perspectives about our results. 

This work is different from the model of sense of belonging proposed by \citet{trinkenreich2023belong}, which investigated the association between intrinsic motivations and the Sense of Virtual Community (SVC). Here we investigate the feeling of welcomeness, which anticipates belonging \cite{hailey2022racialized}, and its association with challenges, not motivations. Moreover, \citet{trinkenreich2023belong}'s study used the diversity lens of gender, while the present study also tackles the lens of race and disabilities.

%The remainder of this section discusses the data analysis and collection.

\vspace{-5pt}
\subsection{Dataset}
\label{sec:data_collection_and_analysis}
\vspace{-5pt}

We used the data from the Linux Foundation Diversity and Inclusion survey \cite{dei_survey}. The survey design was led by the Linux Foundation Research, in collaboration with 40 industrial partners and researchers from our community~\cite{dei_survey}---including an author of this paper, who provided ideas and feedback on the questionnaire. This collaboration between industry and academic researchers helped increase the instrument's clarity, reliability, and reach. The questionnaire was available between July 15 and July 29, 2021, in multiple languages, but with the majority of responses (89\%) in English.% English, Arabic, Chinese (simplified), French, German, Hindi, Japanese, Korean, Portuguese, Russian, and Spanish. The majority of responses (89\%) were in English.

The survey was advertised via social media and through the channels of the project partners (e.g., Linux Foundation, Amazon, CHAOSS Community, Comcast, Fujitsu, GitHub, GitLab, Hitachi, Huawei, Intel, NEC, Panasonic, Red Hat, Renesas, VMware, etc.). The survey was targeted at open source contributors. % and was designed. 
The compensation for answering the survey was a 20\% discount when registering to attend the Open Source Summit Conference. The survey received 2,350 responses.

Linux Foundation Research made the questionnaire and the quantitative data publicly available~\cite{dei_survey_data}, and we obtained the qualitative data by contacting Linux Foundation Research.

The survey included a question about the feeling of welcomeness, a 5-point Likert scale item (from Strongly Disagree to Strongly Agree, with neutral). The questions about the frequency of experiencing challenges were asked as 4-point items: Never, Rarely, Occasionally, and Frequently. For the gender and race questions, respondents could mark as many options they identified with. The disabilities question could be answered as \textit{yes} or \textit{no}.
%, with the possibility of identifying it as a long-term physical, mental, intellectual, or sensory impairment.
%which, in interaction with various attitudinal and environmental barriers, hinders the respondent's full and effective participation in society on an equal basis with others. 
%The survey also included questions about different types of OSS participation, and respondents could mark how often they focus their efforts on different types of OSS participation on a 4-point Likert scale~\cite{replication_package}.
Since our analysis techniques require complete responses, we removed responses with missing values, as they represented more than 10\% of responses, and imputation methods could bias results for multi-group analyses used in RQ2 \cite{hair2017updated}. Given the large number of responses, the trade-off between introducing a potential bias versus increasing the usable sample size suggests a more conservative approach, so we did not use imputation methods. We removed the responses with blank values for: the feeling of welcomeness question (n=4); any interpersonal challenge items (n=104); gender (n=30); race (n=1503); and disabilities (n=2). 

Post filtering, our dataset included 706 responses for the analyses. We present demographics in Table \ref{tab:demographics}.

\vspace{-5px}
\begin{table}[htb]
\caption{Demographics of respondents (n=707)}
\vspace{-5px}
\centering
\label{tab:demographics}
\resizebox{0.7\columnwidth}{!}{
\begin{tabular}{lrr}
\hline
\textbf{Attribute} & \textbf{N} & \textbf{Percentage} \\ 
\hline
\rowcolor{gray!30} \multicolumn{3}{c}{Gender} \\
\textbf{Only Men} & \textbf{539} & \textbf{76.3\%} \\
\textbf{Gender Minorities} & \textbf{167} & \textbf{23.7\%} \\
\quad Women & 115 & 16.3\% \\
\quad Non-Binary or Third Gender~~~ & 29 & 4.1\% \\
\quad More than one gender & 23 & 3.3\% \\
\hline
\rowcolor{gray!30} \multicolumn{3}{c}{Disability} \\
\textbf{No} & \textbf{549} & \textbf{77.8\%} \\
\textbf{Yes} & \textbf{157} & \textbf{22.2\%} \\
\hline
\rowcolor{gray!30} \multicolumn{3}{c}{Race} \\
\textbf{Only White} & \textbf{486} & \textbf{68.8\%} \\
\textbf{Racial Minorities} & \textbf{220} & \textbf{31.2\%} \\
\quad Black & 50 & 7.1\% \\
\quad Hispanic & 43 & 6.1\% \\
\quad Asian & 43 & 6.1\% \\
\quad Native or Indigenous & 4 & 0.6\% \\
\quad Pacific Islander & 2 & 0.3\% \\
\quad More than one race & 78 & 11.0\% \\
\hline
\end{tabular}}
\end{table}
\vspace{-3pt}

We added the subset of questions used in the study in the replication package \cite{replication_package}, together with prepared datasets to run RQ1, RQ2, and RQ3, and the codebook used for qualitative analysis. For the sake of confidentiality between researchers and The Linux Foundation Research, the raw data from the open question could not be shared.
%\footnote{\url{https://data.world/thelinuxfoundation/2021-diversity-equity-and-inclusion-in-open-source}}.

\vspace{-5pt}
\subsection{Hypothesis Development}
\vspace{-5pt}

In the context of this work, \textbf{Feeling Welcomeness} refers to the self-perception of an individual feeling that their presence is desired, accepted, and valued by their OSS colleagues without acts of microaggressions \cite{bembich2017early,stewart2019racial,stewart2019racial}. 
Welcomeness is an inherently emotional and relational concept that entails more than simply being permitted to enter and includes inter-relational subjectivity \cite{gill2018suppression}.
%exists in a world of shared vulnerability and inter-relational subjectivity \cite{gill2018suppression}. 
Understanding the feelings of welcomeness in a given context plays an important role in identifying where barriers exist and how to break down these barriers to increase diversity in participation \cite{bopp2017you}.

Whether an OSS community has a welcoming atmosphere is determined by the challenges that individuals face during the contribution process \cite{balali2018mentor}, affecting both the retention of existing and the attraction of new contributors. 
%While contributors face different types of challenges, the ones that mainly affect attraction and retention are interpersonal-related ones \citep{steinmacher2021being,balali2018newcomers,steinmacher2015social}. 
Interpersonal challenges are difficulties or conflicts that arise from interpersonal relationships (or lack thereof) in OSS projects \cite{lee2019floss,steinmacher2015social}, and tend to have a more pronounced impact on demographics typically underrepresented in the tech industry \cite{lee2019floss,balali2018mentor,trinkenreich2022women}. These challenges can stem from differences in communication styles, cultural backgrounds, personal behaviors, and social dynamics, and can negatively impact collaboration, productivity, and the overall well-being of the individuals involved \cite{steinmacher2015social}.

In the questionnaire, interpersonal challenges were defined based on (1) the unacceptable behaviors listed in the Linux Foundation's code of conduct \cite{lfcodeofconduct} (unwelcoming language, stalking, sexual harassment), (2) issues gathered from the literature and seen by practitioners (lack of response to the rejection of contributions, stereotyping, doxxing), and (3) an overarching conflict-related question (conflict or tension with other contributors).

Given the potential relationship between interpersonal challenges and the feeling of welcomeness, we propose the following hypothesis:

\hypothesis{1}{\textit{Experiencing interpersonal challenges is negatively associated with feeling welcome.}}

We used Partial Least Squares--Structural Equation Modeling (PLS-SEM) \cite{ringle2015structural} to analyze the relationship between interpersonal challenges and feelings of welcomeness in OSS. SEM facilitates the simultaneous analysis of relationships among constructs that can be measured by one or more indicator variables. The main advantage of SEM is being able to measure complex model relationships while accounting for measurement error when using latent variables (e.g., Interpersonal Challenges).

In the following, we discuss the measurement model (i.e., operationalization of constructs) and analysis.

\textbf{Interpersonal challenge} is a theoretical concept that cannot be directly observed given its multiple facets. Therefore, this concept is represented as a latent variable. A latent variable cannot be directly measured but instead is measured through a set of indicators or manifest variables. The interpersonal challenges were measured in our dataset via seven 4-point Likert scale items representing different interpersonal challenges.

1. \textit{"Doxxing"}, derived from "\textit{document dropping}," 
%or "\textit{dropping documents}," 
involves the public disclosure of private or personal information about an individual or organization, often with harmful intentions. 
%This disclosure can include details like real names, addresses, phone numbers, email addresses, and financial records. 
Regarded as a detrimental and unethical action, doxxing can result in privacy breaches and, in some cases, physical harm to the affected parties~\cite{dubberley2020digital}. 
%Often categorized under Open Source Intelligence (OSINT), doxxing is considered a persistent risk. OSINT involves collecting publicly available personally identifiable information about an individual or organization, which is then exploited for various purposes, including harassment, blackmail, hacking, revenge, financial gain, or other motives~\cite{khanna2016experimental}.

2. \textit{Sexual harassment} is a well-documented social issue that can occur offline or online with profound impacts on individuals and organizations. In OSS, conferences and meetups serve as platforms for in-person interaction and can expose 
%. Despite their role in fostering social connections among predominantly online communities, such gatherings, often white-men-dominated, can expose 
minority attendees to inappropriate behavior. 
%Although awareness of this issue is increasing at conferences, there are obstacles to implementing enduring solutions. 
Sexual harassment can occur online also. For instance, when a newcomer who identifies as a woman seeks mentorship from a man, her intentions are often responded to as a dating opportunity \cite{nafus2012patches}.

3. \textit{Stalking} refers to actions meant to intimidate or instill fear \cite{pittaro2007cyber,brady2017dark}, which happens in OSS as a form of toxicity \cite{tourani2017code}.
%Stalking tactics are often widely misconstrued and, in some instances, not clearly defined \cite{brady2017dark}. 
%The literature has found stalking in OSS as a form of toxicity \cite{tourani2017code}.
%, which encompasses a wide range of negative behaviors, including overt insults, rude and disrespectful comments, sarcasm, and microaggressions \cite{cohen2021contextualizing}. In the OSS context, stalking is also addressed through codes of conduct and training \cite{tourani2017code}. 
%Previous studies emphasized kindness as an essential skill to reduce toxicity, collaborate, and be part of an OSS community \cite{liang2022understanding}.

4. \textit{Stereotyping} represents discrimination based on perceived demographic characteristics \cite{mccauley1980stereotyping}, usually referring to a form of negative and fixed impression that can happen implicitly or explicitly and relate to a socially shared set of beliefs about traits of members of a social category~\cite{mccauley1980stereotyping}. 

5. \textit{Unwelcoming Language} is defined as written or spoken communication that induces feelings of exclusion, including instances of profanity, racist jokes, and name-calling. Those who are in minority are more likely to experience unwelcoming language \cite{nafus2012patches,trinkenreich2022women,vasilescu2015perceptions,paul2019expressions,qiu2019signals}. 
%Unwelcoming or hostile discourse contravenes the inclusive policy norms of various online communities, such as StackOverflow \cite{cheriyan2021norm}. Discriminatory language, offensive language, and harsh critiques, which are occasionally encountered in code reviews and mailing lists, can be particularly hurtful to minority groups \cite{paul2019expressions,balali2018newcomers,qiu2019signals}. 

6. \textit{Conflict or tension with other contributors} can manifest in diverse ways and may stem from variations in technical approaches, communication styles, or personal interactions. For instance, tension may arise due to misunderstandings, misinterpretations, or the use of ineffective communication channels \cite{robles2014floss}. Technical debates, such as differing opinions, design philosophies, or coding practices \cite{guizani2021}, can also be sources of tension among contributors.

7. \textit{Lack of response to or rejection of contributions or questions} refers to situations where contributors do not receive acknowledgment or approval for their work or when questions are met with indifference, silence, or dismissal. 
%Contributors can feel neglected when their comments go unanswered, or the response is excessively delayed \cite{miller2022did}. 
Delayed or absent responses lead to contributor attrition in OSS \cite{kaur2022exploring,steinmacher2013newcomers,miller2022did}

%\vspace{-5pt}
\subsection{RQ1 Analysis}
\vspace{-5pt}

To answer RQ1, we quantitatively analyzed the responses to the questions about interpersonal challenges and the feeling of welcomeness using Structural Equation Models--Partial Least Squares (PLS-SEM), utilizing SmartPLS version 4. The first step in evaluating a structural equation model is to assess the soundness of the measurement of the latent variable---this is referred to as evaluating the `measurement model' \cite{hair2019use}. The second step is to evaluate the theoretical model, which includes the evaluation of the hypothesis.
The measurement model is evaluated through tests of convergent validity, internal consistency reliability, discriminant validity, and collinearity. The theoretical model is evaluated according to the significance of the path coefficient and model fit, which are obtained through a `bootstrapping' procedure: it draws a large number (e.g., 5,000) of random `subsamples' of the same size as the original sample (using replacement). The model is estimated for each subsample, generating a sampling distribution used to determine a standard error \cite{hair2019use} that can subsequently be used to make statistical inferences. The model fit is then evaluated using Standardized Root Mean Square Residual (SRMR).
%So, the mean path coefficient determined by bootstrapping can differ slightly from the path coefficient calculated directly from the sample.

%\subsubsection{Validating the Measurement Model}
%\label{sec:results_measurement_model}

\vspace{-1pt}
\subsection{RQ2 Analysis}
\vspace{-3pt}

To answer RQ2, we evaluated the impact of interpersonal challenges (as a single construct) on the feeling of welcomeness across demographic groups. 
To define the demographic groups of interest, we followed the core categories identified by Albusays et al. \cite{albusays2021diversity}. Gender and race are very relevant perspectives in OSS, with many communities implementing initiatives specifically focused on these populations, whereas disability remains underexplored, likely due to data scarcity \cite{silveira2019systematic}. By focusing on these categories, we sought relevance, maintained a clear focus on the paper, and avoided data dredging (i.e., indiscriminately including all data).

We employed Multi-Group Analysis (MGA) to understand how the model varies within distinct demographic groups. We followed \citet{hair2017advanced} proposed structure that includes testing for invariance and comparing groups to analyze results.
%In this stage, we used the ``feeling of welcomeness" variable as originally answered in ordinal (5-point Likert-scale item).

%\textit{Step 1. Group Creation.}
%We grouped our participants to observe any potential heterogeneity among the demographic groups of: (i) gender minorities and men; (ii) race minorities and white; and (iii) Having or not having disabilities. Gender minorities (those who identify their gender as woman, non-binary or third gender, another gender, and more than one gender), non-White (those who identify their race as either Black, Asian, Native Indigenous, Pacific Islander, Hispanic/LatinX, and more than one race), and (dis)abilities (see Table \ref{tab:demographics}).

\textit{Testing for Measurement Invariance of Composite Models (MICOM)} is a mechanism to assess whether or not the loadings of the items representing the latent variables differ significantly across different groups. In other words, we want to establish whether any differences can be attributed to the theoretical constructs and not how we measured those constructs \cite{hair2017advanced}. The literature recommends conducting MICOM before MGA to confirm its appropriateness \cite{hair2017advanced}. If we find differences in the measurement model for the two groups, then we cannot attribute any differences in the model to the theoretical constructs. 

Comparing group-specific model relationships for significant differences using a multi-group analysis requires establishing configural and compositional invariance \cite{henseler2016testing,hair2017advanced}. Configural invariance does not include a test and is a qualitative assessment to ensure all of the composites are equally defined (``configured") for all of the groups, such as equivalent indicators per measurement model, equivalent treatment of the data, and equivalent algorithm settings or optimization criteria. Compositional invariance exists when the composite scores are the same across both groups; while small differences will naturally happen for different groups, we tested whether those differences are significantly different. We used MICOM procedure to examine the correlation between the composite scores of groups and assess that the correlation equals 1.

\textit{Groups Comparison and Analysis} was done through bootstrapping multigroup analysis, and the statistical differences are assessed through parametric tests between path coefficients generated from different samples.
%We analyzed the differences between the coefficients' paths for the groups. If they are significant, they can be interpreted as having moderating effects.

% \subsubsection{RQ2(B) stage}
% \label{sec:method_rq2b}

%(see Section \ref{sec:results_rq2} - Step 3).
%To have more precise information about (i) whether each demographic group experiences each interpersonal challenge and (ii) which interpersonal challenge impacts the feeling of welcomeness of each demographic group, for the second stage, we used a mixed quantitative analysis. 
%We started by analyzing (i) the likelihood of each interpersonal challenge to be experienced by each demographic group. Following, 

\vspace{-2pt}
\subsection{RQ3 Analysis}
\vspace{-3pt}

Results from RQ1 showed that interpersonal challenges are associated with less feeling of welcome, and RQ2 showed that this association is more prevalent for gender minorities and people with disabilities. However, much is not known about the frequency of challenges each demographic experiences. 
%all gender and race minorities are less likely to feel welcomeness than their counterparts, which aligns with previous research \cite{trinkenreich2023belong,trinkenreich2024unraveling} that showed gender minorities feel less belonging than men.
%, being ``an outsider by virtue of [their] gender", ``not treated as a valuable contributor and not feeling psychologically safe to work in [the OSS] space," which aligns with previous research \cite{trinkenreich2023belong} that showed gender minorities still feel less belonging than men.
%The frequency of interpersonal challenges eventually contributes to their feeling of welcomeness. 
%Therefore, we investigated the frequency of experiencing each of the interpersonal challenges for each demographic group. Understanding the feeling of welcomeness, we investigate which demographic contributors are more likely to face which type of challenge. 

To answer RQ3, we built seven individual logistic regression models, with each interpersonal challenge as the dependent variable for each model. The independent binary variables are gender minorities, race minorities, and disabilities (Table \ref{tab:demographics}), allowing us to assess which group will more likely encounter each interpersonal challenge.
%---by understanding the frequency of each challenge. 
We used ordinal logistic regression since the dependent variables were from a 4-point Likert-scale response (never, rarely, occasionally, and frequently). The ordinal logistic regression is an extension of the logistic regression that can be used when the dependent variable is ordinal \cite{harrell2015ordinal}.

\vspace{-2pt}
\subsection{Composite Assessment}
\vspace{-3pt}

We qualitatively analyzed the open-ended responses to the question about the reasons for (not) feeling welcome by inductively applying open and axial coding. 
%We identified the reason each participant provided. 
We built post-formed codes as the analysis progressed and associated them with respective parts of the answer to code the reasons according to the respondents' perspectives. Reasons to not feel welcome were related to six out of the seven interpersonal challenges, and one technical challenge. We also found a positive reason to feel welcome.
We used a negotiated agreement protocol and discussed the codes and categorization as a team until reaching a consensus about the themes, as cataloged in our codebook that is part of our replication package \cite{replication_package}.
%The researchers analyzed each answer and derived codes according to the content of each response. For simplicity's sake, we decided to adopt the same theme nomenclature from the seven interpersonal challenges asked in closed questions (see Sec. \ref{sec:theory_development}), and similar studies about challenges faced by OSS contributors \cite{steinmacher2021being,balali2018newcomers,steinmacher2015social}.
%The outcome was a set of higher-level categories, as cataloged in our codebook~\footnote{https://figshare.com/xxxx}. 

%We classified the codes into four categories defined and used in previous studies \cite{steinmacher2021being,balali2018mentor,steinmacher2015social}: \textsc{Interpersonal} (related to the relationship among other contributors, maintainers, or community), \textsc{Process} (imposed by internal procedures or practices), \textsc{Technical} (related to or caused by technology, including frameworks, programming languages, and tools used to contribute), and \textsc{Personal} (related to personal characteristics of contributors). 

From the 706 answers we used for quantitative analysis, we received 588 answers for the open question of reasons to (not) feel welcome, and we identified respondents as P1 to P588.

%Regarding gender, 80.3\% (1,492 out of 1,859) identified as men, 14.4\% (268 out of 1,859) as women, 3.2\% (60 out of 1,859) as non-binary, and 2.1\% (39 out of 1,859) did not inform their gender. Regarding race, 23.3\% (433 out of 1,859) identified as White, 2.7\% (51 out of 1,859) as Black, 2.2\% (40 out of 1,859) as Hispanic, 1.8\% (33 out of 1,859) as Asian, 0.2\% (3 out of 1,859) as Native or Indigenous, 3.8 (70 out of 1,859) with more than one race, and 66.1\% (1,662 out of 1,859) did not inform their race. Regarding disabilities, 83.1\% (1,545 out of 1,859) did not report any disabilities, 16.8\% (312 out of 1,859) reported disabilities, and 0.1\% (2 out of 1,859) did not inform.

%The first author qualitatively analyzed the answers for the open questions by inductively applying open coding\cite{miles1994qualitative} to organize what participants reported. We built post-formed codes, having the authors conduct card sorting sessions~\cite{Spencer2009}, including discussing the codes and categorization until reaching a consensus about the codes according to the participants' perspectives, who were identified as P1 to P111.

%==============================================================%
%==============================================================%
%==============================================================%

\vspace{-2pt}
\subsection{Follow-up: Community Perceptions}
\label{sec:method_followup}
\vspace{-3pt}

As a follow-up, we evaluated the community's perception of our findings. The goal of the follow-up study was to understand to what extent the interpersonal challenges investigated in this study persist in impacting the feeling of welcomeness in OSS in 2024. We conducted an online survey with OSS contributors, which we advertised on social media and the authors' professional network. We conducted the survey in July 2024. We received 28 responses from OSS contributors from all regions (North America, Central/South America, Europe, Asia Pacific, and Africa). We also conducted an audience poll during the keynote at the 2024 SciPy Conference \cite{SciPy2024}, which is dedicated to scientific computing through open-source Python software. We received an additional 25 responses.

In the surveys, we presented the seven interpersonal challenges and asked if they still happen today and to what extent they can influence how a contributor feels welcome in the OSS community. Additionally, we summarized our results into seven statements and asked the respondents' agreement level. 
%We collected demographics (role, region of living, gender, race, and whether they live with any disability). 
Finally, we added an open question so they could share additional thoughts about challenges and feelings of being included in OSS. All questions were optional.
The questionnaire and its answers are part of the replication package~\cite{replication_package}.

\vspace{-2pt}
\section{Analyses and Results}
\label{sec:results}
\vspace{-3pt}

In this section, we assess each research question, followed by a composite assessment of the answers to the open question about the reasons to (not) feel welcome.

\vspace{-2pt}
\subsection{\rqone}
\label{sec:results_rq1}
\vspace{-1pt}

In this section, we describe our results for RQ1, which include the validate the measurement model (Sec.~\ref{sec:results_measurement_model}), and then we evaluate the hypothesis (\textbf{H1}. Experiencing interpersonal challenges is negatively associated with feeling welcome) and control variables in the structural model (Sec.~\ref{sec:results_structural_model}), both computed through the survey data.

We assess the significance of our model using the Structural Model Evaluation - Partial Least Squares (PLS-SEM) and following the evaluation protocol proposed by previous research \cite{hair2019use,russo2021pls} to make results consistent with our claims. The path weighting scheme was estimated using SmartPLS 4 \cite{sarstedt2019partial}.

PLS-SEM is widely used in software engineering research to explore relationships between theoretical concepts not directly observable but measured through indicators or manifest variables \cite{russo2021pls}. In our study, this applies to interpersonal challenges. As in related work \ref{sec:related}, PLS-SEM constructs a ``proxy" of manifest variables as weighted composites and tests their association with another variable—welcomeness.

\subsubsection{Measurement Model Evaluation}
\label{sec:results_measurement_model}

%We also inspected the model's predictive relevance by means of Stone-Geisser’s \textit{Q}\textsuperscript{2} \cite{stone1974cross} value, which is a measure of external validity \cite{hair2019use}. This measure can be obtained through the PLS-Predict procedure (available within the SmartPLS software). PLS-Predict is a holdout sample-based procedure that generates point predictions on both the item level and the construct level, dividing the sample data into \textit{k} subgroups (referred to as folds) of roughly the same size and combining \textit{k}-1 folds into a training sample that is used to estimate the model. The remaining fold serves as a holdout sample that is used to assess the model’s predictive power \cite{hair2021executing}. \textit{Q}\textsuperscript{2} value is calculated only for endogenous variable of inclusiveness, which led to xxx. Values larger than 0 indicate the construct has predictive relevance, while negative values show the model does not perform better than the simple average of the endogenous variable would do.

We conducted a series of tests to validate the measurement of the theoretical concept, including convergent validity, internal consistency reliability, discriminant validity, and collinearity.

%\subsection{Convergent Validity}
Convergent validity relates to the degree to which a measure correlates positively with alternative measures of the same construct. Our model contains one reflective latent variable (Interpersonal Challenges). %which is reflective (not formative) as a function of the latent construct. 
Changes in the latent construct should be reflected in changes in the indicator variables \cite{hair2019use}. We used two metrics to assess convergent validity: the Average Variance Extracted (AVE) and the loading of an indicator onto its construct (the outer loading) \cite{hair2019use}. The AVE value for the latent construct in our model is 0.52, which is above the threshold of explainability (0.5). 
For the outer loading, we remove variables that does not reflect a sufficiently large impact in the latent variable. An outer loading of 0.708 is considered sufficient (because 0.708\textsuperscript{2} $\approx$ 0.50, which means at least half of the indicator's variance can be explained by the latent variable), and 0.60 is considered sufficient for exploratory studies \cite{hair2019use}. All indicators of our latent construct exceeded this threshold, ranging between 0.62 and 0.83.

%For the outer loading, we remove any variable that does not reflect a sufficiently large impact in the latent variable (outer loading $>0.4$). 

%The AVE is equivalent to a construct's commonality \cite{hair2019use}, i.e., the proportion of variance shared across indicators. A reflective construct is assumed to reflect (or ``cause'') any change in its indicators. The AVE should be at least 0.5, indicating that it explains most of the variation (i.e., 50\% or more) in its indicators \cite{hair2019use}. This variance is indicated by taking the squared value of an indicator's loading. The AVE value for the latent construct in our model is above this threshold (0.52).

%A latent variable is measured by two or more indicators; indicators with loading below 0.4 should be removed because this implies that a change in the latent construct that it purportedly represents (or `reflects') does not get reflected in a sufficiently large change in the indicator \cite{hair2019use}. An outer loading of 0.708 is considered sufficient (because 0.708\textsuperscript{2} $\approx$ 0.50, which means at least half of the indicator's variance can be explained by the latent variable), and 0.60 is considered sufficient for exploratory studies \cite{hair2019use}. All indicators of our latent construct exceeded this threshold, ranging between 0.62 and 0.83.

%\subsection{Internal Consistency Reliability}

We also verified the Internal Consistency Reliability, or how well the different indicators for interpersonal challenges are consistent with one another and able to reliably and consistently measure the constructs. %A high degree of consistency suggests that indicators refer to the same construct. There are several tests to measure internal consistency reliability. 
We calculated both Cronbach's \textalpha{} and Composite Reliability (CR) to measure the reliability~\cite{hair2019use}. %Cronbach's \textalpha{} frequently shows lower values, whereas the Composite Reliability (CR) may overestimate slightly \cite{hair2019use}.
A desirable range of values for both Cronbach's \textalpha{} and CR is between 0.7 and 0.9 \cite{hair2019use}. %Values below 0.6 suggest a lack of internal consistency reliability, whereas values over 0.95 suggest that indicators are too similar and thus are not desirable. x
The Cronbach's \textalpha{} was 0.871, and CR was 0.887 for our model. 
%Further, we also report the consistent (exact) reliability coefficient ($\rho_a$) \footnote{$\rho_a$ values typically lie between Cronbach's $\alpha$ and CR, serving as a good compromise between the measures and addressing the underestimations of Cronbach's \textalpha{} \cite{sarstedt2021partial}}.

%\subsection{Discriminant Validity}

We also verified whether each construct represented different concepts or entities through discriminant validity tests to assess the discriminant validity utilizing the Heterotrait-monotrait (HTMT) ratio of correlations \cite{henseler2015new}. The discriminant validity could be problematic if the HTMT ratio exceeds 0.9 \cite{henseler2015new}. The HTMT ratio between the latent construct and the endogenous variable was 0.460, which is not a significant threat to this study. Finally, we evaluated if the indicators from the exogenous construct (Interpersonal Challenges) were independent. We calculated their collinearity using the Variance Inflation Factor (VIF). A widely accepted cut-off value for the VIF is 5 \cite{hair2019use}; in our model, all VIF values are below 2.

Given these results, we argue that the model is valid for evaluating our hypothesis.

\subsubsection{Evauating the Hypothesis in the Structural Model}
\label{sec:results_structural_model}

\begin{figure}[htb]
\centering
\includegraphics[width=0.5\textwidth]{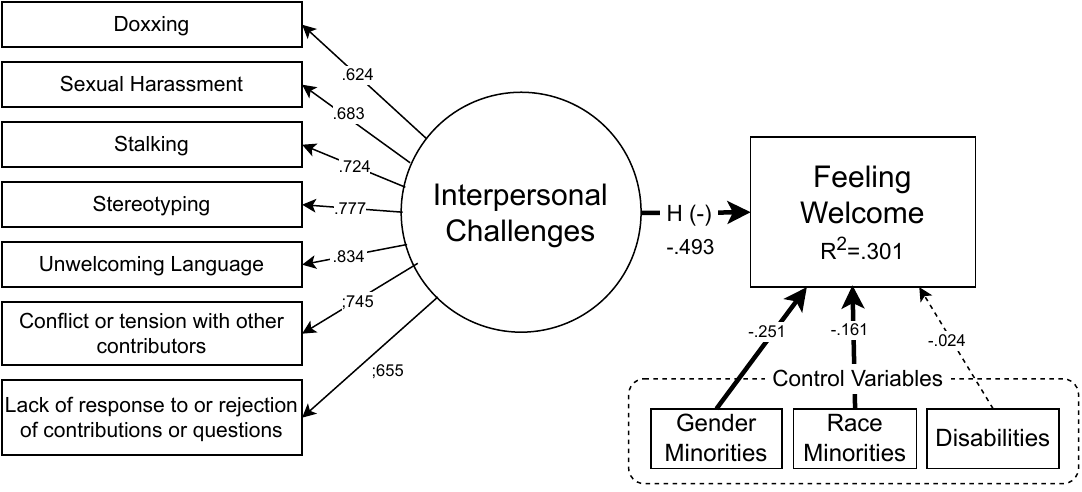}
\caption{\footnotesize Item loadings and path coefficient (p $<$ 0.05 indicated by a full line). Non-significant links are indicated with a dashed line.
} 
\label{fig:evaluating_structutal_model}
\vspace{-10px}
\end{figure}

We found support for our hypothesis (p = 0.000): experiencing interpersonal challenges is negatively associated with feeling welcome. This relationship is depicted by an arrow in the diagram in Fig.~\ref{fig:evaluating_structutal_model}. The path coefficient, interpreted as a standardized regression coefficient, indicates the direct effect of one variable on another. With a negative path coefficient (\textit{B}=-0.493), we conclude that interpersonal challenges are negatively associated with the feeling of being welcome. This means that an increase of one standard deviation in Interpersonal Challenges results in a decrease of 0.493 standard deviations in Feeling Welcome.

%We found support for our hypothesis (p = 0.000): experiencing interpersonal challenges is negatively associated with feeling welcome. This hypothesis is represented by an arrow in the diagram in Figure~\ref{fig:evaluating_structutal_model}. The path coefficient is interpreted as a standardized regression coefficient, indicating the direct effect of one variable on another. Given its negative path coefficient (\textit{B}=-0.493), we conclude that interpersonal challenges are negatively associated with feeling welcome. The path coefficient (\textit{B} is -0.493), which means that when the score for Interpernosal Challenges increases by one standard deviation unit, the score for Feeling Welcome decreases by 0.493 standard deviation unit (the standard deviation is the amount of variation of a set of values). 

We assessed the relationship between constructs and the predictive capabilities of the theoretical model. The \textit{R}\textsuperscript{2} value of the endogenous variable in our model (feeling welcome) was 0.301. While some scholars have suggested thresholds to evaluate such values, there is considerable debate about setting such thresholds. Other factors might play a role in feeling welcome. We consider the \textit{R}\textsuperscript{2} value low-moderate.

We analyzed the Standardized Root Mean Square Residual (SRMR), which is a common fit measure to detect misspecification of PLS-SEM models \cite{russo2021pls}. SRMR is a model fit measure that quantifies the divergence between observed and estimated covariance matrices. The literature suggests using SRMR to compare and evaluate SEM models \cite{henseler2016using}. SRMR is the square root of the sum of the squared differences between the model-implied and the empirical correlation matrix, or the Euclidean distance between the two matrices \cite{henseler2014common}. 
A value of 0 for SRMR would indicate a perfect fit, and values less than 0.08 (conservative) or 0.10 (more lenient) are considered a good fit \cite{henseler2016using}. Our results suggest a good fit of the empirical data in the theoretical model (SRMR = 0.075).

\textit{Control Variables:} We also examined our data to determine if being part of gender or race minorities or having disabilities could strengthen or weaken feeling welcome. We found that identifying as gender or race minorities is associated with not feeling welcome, but we did not find a significant association between having disabilities and feeling welcome.

Table~\ref{tab:path_analysis_new} shows the results of the analysis, including the mean of the bootstrap distribution (\textit{B}), the standard deviation (\textit{SD}), the 95\% confidence interval, and p-values.

\vspace{-5px}
\begin{table}[htb]
\centering
\caption{Standardized path coefficients (\textit{B}), standard deviations (Std), confidence intervals (95\% CI), and p-values.}
\label{tab:path_analysis_new} % Note: I've changed the label to avoid conflicts if you're using both versions in the same document.
\vspace{-5px}
\resizebox{\columnwidth}{!}{
%\resizebox{0.65\textwidth}{!}{
\begin{tabular}{lrrrr}
\toprule
& {\textit{B}} & Std & 95\% CI & p-value \\
\midrule
H1 
%Interpersonal Challenges $\rightarrow$ Feeling Welcome 
& {\cellcolor{gray!20}\textbf{-0.49*}} & 0.035 & (-0.56, -0.42) & $<$0.001 \\
Gender Minorities $\rightarrow$ Feeling Welcome & {\cellcolor{gray!20}\textbf{-0.25*}} & 0.078 & (-0.41, -0.10) & 0.001 \\
Race Minorities $\rightarrow$ Feeling Welcome & {\cellcolor{gray!20}\textbf{-0.16*}} & 0.073 & (-0.31, -0.02) & 0.028 \\
Disabilities $\rightarrow$ Feeling Welcome & -0.02~~ & 0.085 & (-0.19, 0.14) & 0.779 \\
\bottomrule
\end{tabular}}

\resizebox{0.9\linewidth}{!}{%
\begin{tabularx}{\linewidth}{lX}
%\begin{tabularx}{\linewidth}{@{}@{}}
Coefficients marked with * are statistically significant 
\end{tabularx}
}
\vspace{-10px}
\end{table}

% \begin{table}[!t]
% %\footnotesize
% \centering
% \caption{Standarized path coefficients, standard deviations, confidence intervals, and p-values. Coefficients marked with * are statistically significant}
% \vspace{-3mm}
% \label{tab:path_analysis}
% \robustify{\bfseries}
% \sisetup{
%     mode=text,
%     group-digits = false ,
%     input-signs ={-},
%     input-symbols = ( ) [ ] - + *,
%     detect-weight=true, 
%     detect-family=true,    
%     add-decimal-zero=false, %% doesn't seem to work :-(
%     add-integer-zero=false,
%     round-mode=places, 
%     round-precision=2, %% change this for precision.
%     parse-numbers = true
% }
% \footnotesize
% \begin{tabular}{P{4cm}
%                 S[table-format=0.2]
%                 S[table-format=0.2]
%                 C{1.5cm}
%                 S[table-format=0.3,round-precision=3]}
% \toprule
% & {\textit{B}} & {SD} & {95\% CI} & p\\
% \midrule

% \hangindent1em H1  Interpersonal Challenges $\rightarrow$ Feeling Welcome & -.49* & .035 & (-.56, -.42) & .000\\
% \midrule
% \hangindent1em Gender Minorities $\rightarrow$ Feeling Welcome & -.25* & .078 & (-.41, -.10) & .001\\
% \hangindent1em Race Minorities $\rightarrow$ Feeling Welcome & -.16* & .073 & (-.31, -.02) & .028\\
% \hangindent1em Disabilities $\rightarrow$ Feeling Welcome & -.02 & .085 & (-.19, .14) & .779\\

% \bottomrule

% \end{tabular}
% \vspace{-3mm}\end{table}

\MyBox{\textbf{RQ1.} Experiencing interpersonal challenges is associated with not feeling welcome within OSS projects. Gender and racial minorities feel less welcome than their counterparts.}

\vspace{-7pt}
\subsection{\rqtwo}
\label{sec:results_rq2}
\vspace{-5px}

We answered RQ2 by performing a Multi-Group Analysis in PLS-SEM to investigate the moderation of the theoretical model across demographic groups. We used MGA to examine how the results from RQ1 vary across cohorts.

We started by using the groups of (i) gender, (ii) race, and (iii) disabilities (see Table \ref{tab:demographics}).
Following, we observed that configural invariance is established in our model as no different settings or treatments were applied to the groups. Still, in this step, we ran the permutation test in SmartPLS and verified that compositional invariance is established for all latent variables in the PLS path model. We established partial measurement invariance, and MGA is suitable for the examined groups \cite{ringle2016gain}.

Finally, we analyzed the differences between the coefficients' paths for the groups. If they are significant, they can be interpreted as having moderating effects. In this step, we used parametric tests to assess the multi-group differences, as in Step 2 we established partial measurement invariance \cite{ringle2016gain}. As Table~\ref{tab:mga} shows, parametric tests showed statistical differences regarding gender and disabilities but not regarding race. Gender minorities had (\textit{B}=\num{-0.55}) a 1.6x stronger association between interpersonal challenges and not feeling welcome than men (\textit{B}=0.35). Parametric tests showed no significant differences for race or disabilities.

\begin{table}[!ht]
\centering
\vspace{-2mm}
\caption{Multi-Group Analysis per demographics}
\vspace{-5px}
% \vspace{-2mm}
\label{tab:mga}
%\robustify{\bfseries}
\sisetup{
    mode=text,
    group-digits = false,
    input-signs ={-},
    input-symbols = ( ) [ ] - + *,
    detect-weight=true,
    detect-family=true,
    table-format=0.2,
    add-decimal-zero=false, %% doesn't seem to work :-(
    add-integer-zero=false,
    round-mode=places,
    round-precision=2, %% change this for precision.
    parse-numbers = true
}
\footnotesize
\resizebox{\columnwidth}{!}{%
%\resizebox{0.8\textwidth}{!}{
\begin{tabular}{@{}l
                S[table-format=0.2]
                S[table-format=0.2]
                S[table-format=0.2]
                S[table-format=0.2]
                S[table-format=0.2]
                S[table-format=0.2]
                S[table-format=0.2]
                S[table-format=0.2]@{}}

\toprule
& \multicolumn{2}{c}{Gender} & \multicolumn{2}{c}{Race} & \multicolumn{2}{c}{Disabilities} & {All} \\
\cmidrule(lr){2-3} \cmidrule(lr){4-5} \cmidrule(lr){6-7} \cmidrule(lr){8-8}
Group & {Minorities} & {Men} &  {Minorities} & {White} & {Yes} & {No} & \\
\midrule
Sample size (N) & {167} & {539} & {220} & {486} & {547} & {157} & {706}\\

\midrule
Feeling Welcome (\textit{R}\textsuperscript{2}) & 0.31 & 0.12 & 0.24 & 0.31 & 0.27 & 0.16 & 0.31\\

\midrule
H1
& {\cellcolor{gray!20}\textbf{-0.55*}} & {\cellcolor{gray!20}\textbf{-0.35*}} & {-0.51*} & {-0.55*} & {\cellcolor{gray!20}\textbf{-0.52*}} & {\cellcolor{gray!20}\textbf{-0.40*}} & {-0.49*}\\

\bottomrule

\end{tabular}

}

\resizebox{0.7\columnwidth}{!}{%
\begin{tabularx}{\columnwidth}{X}
%\begin{tabularx}{\linewidth}{@{}@{}}
Coefficients marked with $*$ are statistically significant, and in gray are statistically significant differences between groups
\end{tabularx}
}

%\vspace{-1mm}
\vspace{-10px}
\end{table}

\MyBox{\textbf{RQ2} Interpersonal challenges were more strongly associated with the lack of feeling welcome for gender minorities and people with disabilities, while no significant difference was observed for race minorities in this context.}

\subsection{\rqthree}

\begin{figure*}[!tp]
\centering
\includegraphics[width=7 in]{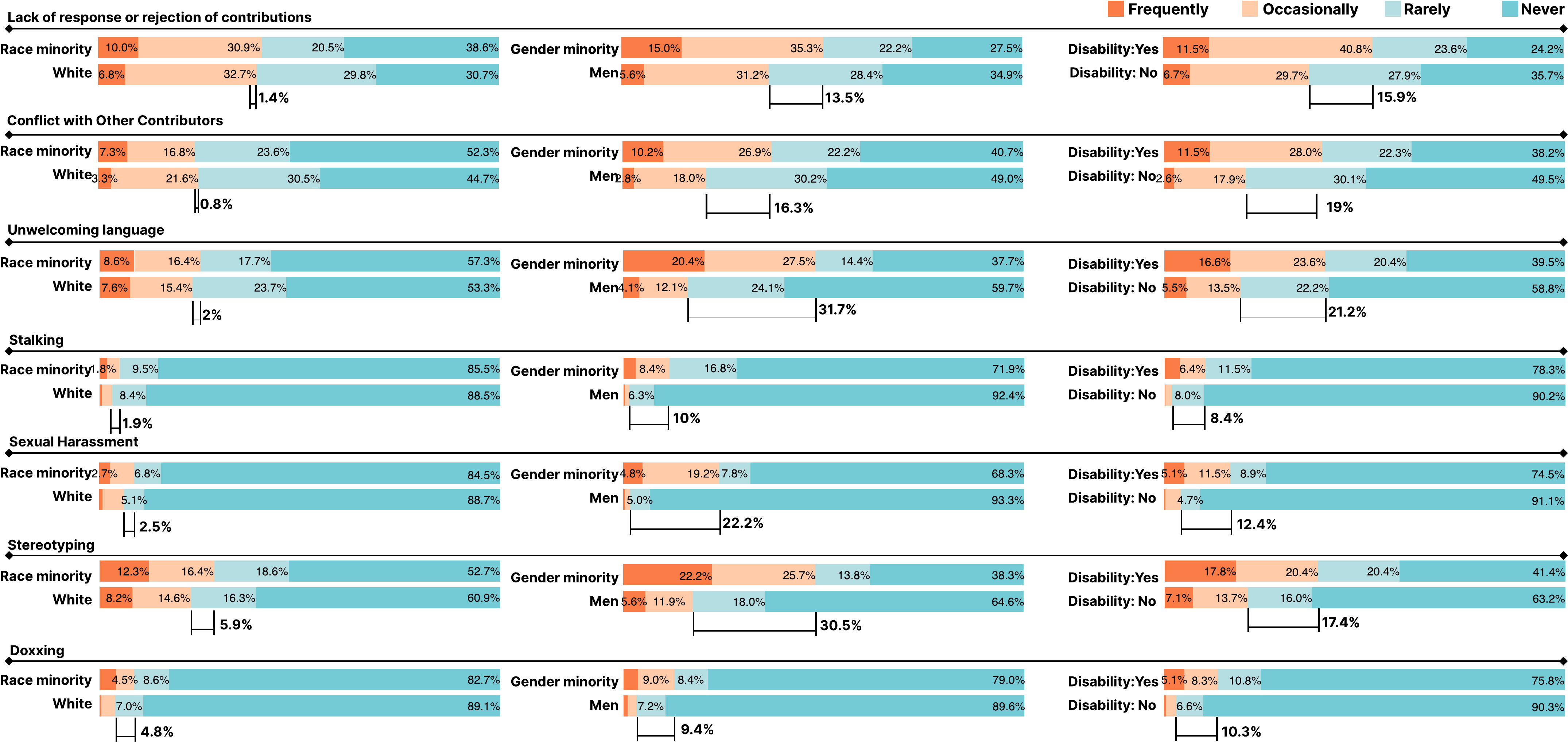}
\caption{\footnotesize Responses to the 4-points Likert-scale items for ``frequency of interpersonal challenges'' (Left:Race; Middle: Gender; Right: Disabilities). The percentage at the bottom of each pair represents the delta, comparing the answers to frequently and occasionally. The whiskers present the differences experiencing challenges (frequent + occasional) between the minority and majority groups.}
\label{fig:freq_challenges}
\vspace{-15px}
\end{figure*}

In this section, we investigate the frequency of interpersonal challenges experienced by the three demographic groups of gender, race, and disability. In Fig.~\ref{fig:freq_challenges}, we present the answers to the questions about the frequency of experiencing interpersonal challenges in OSS (from left to right: Frequently, Occasionally, Rarely, and Never). We analyzed the graph by comparing the percentage of answers in the two top frequencies (Frequently and Occasionally). After each item, we show the delta, comparing the percentages for each minority and its counterpart.

%Our findings indicate different interpersonal challenges experienced by different demographic groups and individuals identifying as gender, race, and disability minorities within the LF community.

Next, we conducted an ordinal logistic regression to follow up on the differences observed in the visualization (Table \ref{tab:logistic}). We have checked our variables' covariance matrix and VIF to confirm no multicollinearity. We used the Benjamini-Hochberg approach to adjust the p-value for the coefficients \cite{thissen2002quick} to control the false rate due to multiple comparisons. 
As we analyzed interpersonal challenges using three demographic variables as independent predictors, resulting in multiple comparisons, we applied the Benjamini-Hochberg approach to adjust the p-values in our ordinal logistic models to control the false discovery rate (FDR) given the multiple hypothesis tests. This method allows control of FDR while maintaining statistical power and without being overly conservative.

An odds ratio $>$ 1 suggests higher odds for minorities to face the challenge more frequently than their counterparts.

\vspace{-8px}
\begin{table}[htb]
\centering
\caption{Odds ratios of challenges per demographics}
\label{tab:logistic}
\vspace{-5px}
\resizebox{\columnwidth}{!}{
\footnotesize
\begin{tabular}{
>{\columncolor[HTML]{FFFFFF}}l 
>{\columncolor[HTML]{FFFFFF}}l 
>{\columncolor[HTML]{EFEFEF}}l 
>{\columncolor[HTML]{EFEFEF}}l }
\hline
\multicolumn{1}{c}{\cellcolor[HTML]{FFFFFF}\textbf{Challenges}} & \multicolumn{1}{c}{\cellcolor[HTML]{FFFFFF}\textbf{Race Minorities}} & \multicolumn{1}{c}{\cellcolor[HTML]{FFFFFF}\textbf{Gender Minorities}} & \multicolumn{1}{c}{\cellcolor[HTML]{FFFFFF}\textbf{Disability}} \\ \hline
Lack of Response or Rejection of Contributions                  & 0.87                                                                 & {\color[HTML]{0D0D0D} \textbf{1.57*}}                                 & {\color[HTML]{0D0D0D} \textbf{1.68*}}                          \\
Conflict with Other Contributors                                & 0.79                                                                 & {\color[HTML]{0D0D0D} \textbf{1.58*}}                                 & {\color[HTML]{0D0D0D} \textbf{1.95*}}                          \\
Unwelcoming Language                                            & 0.85                                                                 & {\color[HTML]{0D0D0D} \textbf{3.09*}}                                 & {\color[HTML]{0D0D0D} \textbf{2.14*}}                          \\
Stalking                                                        & 1.21                                                                 & {\color[HTML]{0D0D0D} \textbf{4.36*}}                                 & {\color[HTML]{0D0D0D} \textbf{1.97*}}                          \\
Sexual harassment                                               & 1.34                                                                 & {\color[HTML]{0D0D0D} \textbf{6.17*}}                                 & {\color[HTML]{0D0D0D} \textbf{2.67*}}                          \\
Stereotyping                                                    & 1.38                                                                 & {\color[HTML]{0D0D0D} \textbf{3.25*}}                                 & {\color[HTML]{0D0D0D} \textbf{1.97*}}                          \\
Doxxing                                                         & 1.67                                                                 & {\color[HTML]{0D0D0D} \textbf{2.03*}}                                 & {\color[HTML]{0D0D0D} \textbf{2.67*}}                          \\ \hline
\end{tabular}}
\resizebox{0.7\columnwidth}{!}{%
\begin{tabularx}{\columnwidth}{X}
%\begin{tabularx}{\linewidth}{@{}@{}}
Coefficients marked with $*$ are statistically significant
\end{tabularx}
}
\vspace{-10px}
\end{table}

Gender minorities and people with disabilities from our sample were more likely to experience all interpersonal challenges than their counterparts (see Table~\ref{tab:logistic} and Fig.~\ref{fig:freq_challenges}). For example, gender minorities were 3.09 times more likely to experience \textsc{unwelcoming language} (with a delta of 31.7\% for the highest frequency answers) and 6.17 times more likely to experience \textsc{sexual harassment} (with delta of 22.5\%) than men. People with disabilities reported 2.14 more odds of experiencing \textsc{unwelcoming language} and 2.67 more odds of experiencing \textsc{doxxing} than their counterparts. 

However, when we look at race minorities, we could not find a statistically significant difference in the odds of experiencing any interpersonal challenge compared to those who identified as white only. %However, this should not conclude the interpersonal challenges that race minorities experience, as the quantitative metrics may always capture the daily activities and experiences. 
From the visualizations (Fig.~\ref{fig:freq_challenges}, we can see that the frequency for all the categories does not vary significantly for any of the items and that for the \textsc{conflict or tension with other contributors} challenge, the number of high-frequency answers is slightly higher for the non-white sample.

%40\% of contributions, irrespective of demographic groups, are frequently facing the challenge of lack of response and reaction of contributions. Similarly, for conflicting with others and unwelcoming language, more than 20\% contributors from both race minorities and whites are equally experiencing the challenge. 

From a cross-demographics perspective, we observed in Fig.~\ref{fig:freq_challenges} that the number of respondents who reported \textsc{lack of response or rejection of contributions} \textit{frequently} or \textit{occasionally} is high cross all demographics, noted by around half of the respondents. Even with the prevalence of this challenge, gender minorities and people with disabilities reported a higher frequency of occurrence (gender minorities 57\% more likely to face the challenge more often than their counterparts, and people with disabilities 68\% more likely).
%This could be attributed to maintainer burnout—a phenomenon widely recognized in existing literature [citation needed]. 

Another major concern for contributors is the \textsc{conflict or tension with other contributors}. While we did not observe any differences regarding race, gender, minorities and people with disabilities reported higher odds (1.58 and 1.95, respectively) of experiencing this challenge than their counterparts. This suggests that these two groups could experience unique challenges related to conflicts. 
We noticed that challenges related to \textsc{unwelcoming language} and \textsc{stereotyping} followed a similar trend. as conflicts. %as there are no differences between race minorities and whites, but for gender minorities, contributors with disabilities always have higher differences than their counterparts. 
To further investigate this, we conducted a follow-up analysis to assess the correlation of \textsc{conflict or tension with other contributors} with \textsc{unwelcoming language} and with \textsc{stereotyping} for the minority groups. We noted that the correlations were statistically significant, with (a least) moderate correlation (Spearman's rank correlation coefficient $\rho$ \cite{sedgwick2014spearman})
---0.71, 0.67, 0.69 for \textsc{unwelcoming language}/\textsc{conflict} and 0.59, 0.59, 0.64 for \textsc{stereotyping}/\textsc{conflict}, considering gender, race, and disabilities minorities, respectively. This may suggest that the conflicts may have a relationship with these other two challenges, especially for gender minorities and contributors with disabilities.

Both Fig.~\ref{fig:freq_challenges} and Table~\ref{tab:logistic} highlight alarming concerns regarding \textsc{sexual harassment}, \textsc{stalking}, and \textsc{doxxing}, which present the highest odds ratio disadvantaging minorities. Even though these interpersonal challenges are less common than others, they bring a non-negligible delta (when we compare the highest frequency answers from gender and disabilities minorities with their counterparts), and a \textit{really low} number of ``frequently" and ``occasionally" answers of the majorities. By looking at Table ~\ref{tab:logistic}, we observe that gender minorities are 6.17 times more likely to experience sexual harassment than men, with high odds ratios also for \textsc{Stalking} (4.36) and \textsc{Doxxing} (2.03); these interpersonal challenges are at least two times more likely to be experienced by gender minorities than by men.

% In addressing RQ3, our findings indicate a differential experience of interpersonal challenges among individuals identifying with gender, race, and disability minorities within the LF community. 

% Needs eyes here took a stab. 

% Stereotyping --> sexual harrassment --> doxxing slippery slope leading to more and more toxicity. 

% RQ3: How differently do those in gender, race, or disability
% minority face interpersonal challenges?

% Here is a figure that shows the frequency of interpersonal challenges.

% Look at lack of response, this is a concern on every aspect. this could be arisiing coz of maintain burnout effects all kinds of majorities and minorities.

% 70\% people have that. Pointed out by literature [].

% next, Look at conflict with other contributors. Gender and disabilities quite a bit of difference although no significant for race, saying <>

% Correlation of conflict with unwelcoming language high. Unwelcoming language also had these differences.

% Unwelcoming language also had strong correlation with stereotyping. Stereo also had big differences for gender minorities and people with disabilities. These known things are talked about [...] but OSS facing bigger concerns. For example, look at Sexual harrasment quite a bit difference. This is actually a big concerns with security and physical well being and stalking and doxxing are also highly correlated. It is not happening much but it does effect women and disabled people. It is a serious concern that OSS community managers can really help. COC can help and can take really strong action. 

\MyBox{\textbf{RQ3} Gender minorities and people with disabilities face significantly more interpersonal challenges compared to their counterparts. The lack of response or rejection of contribution is prevalent among both minorities and majorities, while sexual harassment, stalking, and doxxing are prevalent among minorities. Conflict or tension with other contributors is experienced together with either unwelcoming language or stereotyping by minorities of gender, race, and disabilities.}

\vspace{-5px}
\subsection{Composite Assessment}
\label{sec:results_rq4}
\vspace{-2px}

As part of our mixed method concurrent embedded strategy \cite{creswell2017research}, we had a composite assessment from the open question of reasons to (not) feel welcome. During this strategy, we qualitatively analyzed data to reside side by side with quantitative analysis performed to answer RQ1, RQ2, and RQ3, having different pictures to provide an overall composite assessment of the problem, releaving details about why our participants feel (or do not feel) welcome in OSS. 

%Although we were interested in interpersonal challenges to triangulate the findings from previous RQs, we found process, technical, and personal reasons for (not) feeling welcome. Hence, we organized the reasons using the four categories of interpersonal challenges faced by OSS contributors \cite{balali2018mentor}, including reasons related to \textsc{Process}, \textsc{Technical}, \textsc{Interpersonal} and \textsc{Personal} aspects of the contribution.

%, as presented in Fig. \ref{fig:reasons}.

%However, most of the reasons for not feeling welcome were interpersonal. even answering the questions about interpersonal challenges after the open question about feeling welcome (we discuss this finding in Sec. \ref{sec:discussion}. 
%Table \ref{tab:codes} presents the number of participants whose responses fit each category.

%\input{Tables/codes}

%\begin{figure*}[htb]
%\centering
%\includegraphics[width=1\textwidth]{Figures/reasons.png}
%\caption{Reasons for (not) feeling welcome in OSS. Reasons marked with (*) were investigated in RQ1 and RQ2.} 
%\label{fig:reasons}
%\end{figure*}

%In the following, we present more details about our findings organized by reasons category.

%\subsubsection{Interpersonal-related reasons} are related to the relationship among other contributors, maintainers, or the community \citep{balali2018newcomers} and were mentioned by 48.6\% (54 out of 111) of the respondents.
%\subsubsection{Interpersonal-related Reasons for (Not) Feeling Welcome.}

People are often biased against others outside of their social group, showing prejudice (emotional bias), \textsc{stereotypes} (cognitive bias), and discrimination (behavioral bias). Both minorities and majorities mentioned being discriminated against due to their demographic characteristics of race, disabilities, gender, age, religion, political ideas, or sexual preferences. 
%A \textit{``a strong machismo"} (P1261) is perceived by gender minorities as undermining the value of their contributions. 
For example, race minorities feel discriminated against during in-person gatherings. P80 mentioned being \textit{``always the only African American and people act like: why are you here?"}. During technical group discussion meetings, P141 remembered that \textit{``being African American and the white people would all sit on the other side of the room."} 
%Discrimination was also perceived against Native Americans when trying to get hired in OSS for a job, as \textit{``white people directly hire white people here"} (P1427). 
Discrimination against gender minorities was mostly mentioned as a glass ceiling, as a woman reported \textit{``they are still excluded from the critical discussions."} (P508). 
%The consequence of discrimination is even more pervasive for people who are part of multiple minority groups, older non-binaries who need to ``prove-it again", constantly \textit{``working to defuse assumptions about myself based on [their] age and gender"} (P1596). 
Discrimination against majorities appeared as another side of the spectrum. In a notable contrast, men, white, middle-aged, or respondents who have a straight sexual orientation feel \textit{``forced into a box} (P123), discriminated against and hated because they are not part of minorities. A white man mentioned \textit{``in the last few years, there has been a dramatic increase in the dismissal of straight, white, male contributions and demonizing of the same group."} (P431). 

Contributors mentioned feeling welcome because they can \textit{``hide behind a profile picture"} and be anonymous  (P75, P75, P206, P567). However, when their identities are revealed (\textsc{doxxing}), P551 complemented: \textit{you're dead in the water"}.

Although majorities feel discriminated against, diversity incidents against minorities are still pervasive with \textit{``tons of microaggressions, macroaggressions and \textsc{sexual harassment}"} (P232). Microaggressions were mentioned as other kinds of \textsc{conflicts or tension with other contributors} who can make \textit{``harsh criticism and impersonal responses"} (P71). During conflicts, P8 misses allies because \textit{``when treated poorly, [she does] not have a single memory of another person in a thread stepping up to disagree"}.

Conflicts were mostly mentioned during communication, and respondents reflected on toxic communication during technical discussions, attributing it to the perception of \textsc{unwelcoming language} as being called as \textit{tranny or a dumb faggot"} (P8), which leads to unsafety to ask questions (P54), as when asking a question, they faced hostility in which \textit{``people are too negative when they should be inquisitive or just technical"} (P133).

Besides toxicity, contributors feel unwelcome due to \textsc{lack of response to or rejection of contributions or questions} when they feel maintainers ignore their attempts of making contact (P105, P168), bringing the perception that \textit{``pull-requests will never get merged"} (P526).

%We did not find any explanation related to \textsc{Stalking} as reasons for (not) feeling welcome in the open question. We found that respondents mentioned not feeling welcome due to \textsc{low English confidence}. For example, P1839 said that {``there is a lack of a solid system of translations to bridge the gap for those not comfortable with English"}; still, P78 mentioned that \textit{``English is dominate"} (P78) and several respondents reported that they cannot \textit{``keep up with English conversations"} (P10). \textit{``Although having problems communicating in English"}, a woman reasoned her feeling welcome by \textit{``being able to participate in Spanish translation, which has served the community and has been helpful"} (P1354). Conversely, in general, English confident speakers felt welcome because of being part of majority groups and their fluency in English (e.g. \textit{``as a white cis English-speaking man I do not see many barriers."}
On the positive side, our respondents mentioned feeling welcome due to \textsc{exchanging support} when \textit{``people are helpful and share their knowledge"} (P128), but P148 they pondered there is an \textit{``insular culture"} where they need to \textit{``know insiders"} to feel welcome. P282 agreed that \textit{``in the communities that know [him], [he] feels very welcome.. in areas where [he is]  unknown, [he] feels like [he does] not have a voice/need to prove [him]self."}

Although we were interested in interpersonal-related reasons for (not) feeling welcome, another type of reason appeared from the answers. Respondents mentioned feeling unwelcome due to \textsc{lack of accessible tools}, particularly for people with sensory impairments like blindness. P518 reported missing accessible Linux mobile phones and remembered an episode when they \textit{``asked a PinePhone developer on Mastodon if accessibility was considered, but [were told that] Linux Desktop accessibility had to develop further, and that [they] should be the change [they] want to see in the world. [They] felt very unwelcome, since [they] know [they] can never come to the knowledge needed to fix an accessibility stack"}.

%Still, on technical reasons, we found people saying they felt welcome because they \textsc{have code abilities}, which was often called a \textit{``technical talent"} (P47) or \textit{``competency"} (P114). However, \textit{``being a non-developer role tends to put [them] in a class of people who [are] treated lesser than folks who code."} (P1), as also stated by P463: \textit{``I'm not a developer, and it feels very developer/tech-centric."}

%Finally, \textsc{perceived meritocracy} (P384) in the code review process was also perceived as a reason to feel welcome, which means when \textit{``the code is good, it is accepted"} (P412), so their \textit{``contributions are weighed on their merits"} (P130). Although meritocracy was a reason to feel welcome for mostly majority groups of gender and race, the perception that \textit{``no one cares about anything other than how good the code you write is"} (P104) was mentioned by a non-binary contributor, which shows that code review bias is not universally happening in all projects.

\vspace{-5px}
\subsection{Follow-up: Community Perceptions}
\label{sec:results_followup}
\vspace{-2px}
%We used open coding to qualitatively analyze the 25 answers to the open question about which challenges participants feel still persist in 2024. 

The follow-up survey was designed to provide insights into how the community perceives our findings (see Sec. \ref{sec:method_followup}). We first presented the seven interpersonal challenges and asked to what extent these challenges influence the feeling of welcomeness. Out of the 53 respondents, 39 answered the question. Among them, 89\% reported that the seven interpersonal challenges we investigated continue to significantly or extremely affect welcomeness in OSS. None of the respondents believed these challenges had any effect, indicating that the issues identified in the 2021 survey persisted in 2024 when we conducted the follow-up survey. 
%For the survey respondents, when asked about each challenge separately, most respondents considered that the challenges still happen in 2024; with only one respondent reported that they believed one of the challenges was not happening anymore.  

\begin{figure*}
    \centering
    \includegraphics[width=\linewidth]{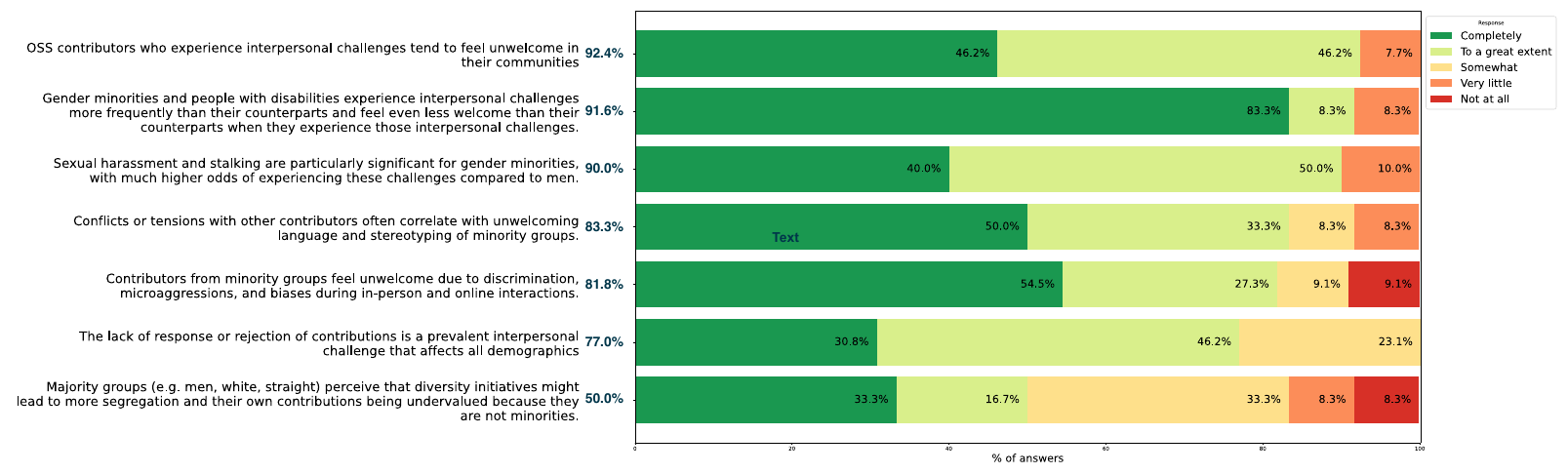}
    \vspace{-5mm}
    \caption{\footnotesize Results from the follow-up survey. Practitioners answered ``How much do you believe these statements hold true in OSS in 2024?'' for statements derived from our findings. The list is sorted by the number of respondents who strongly support the statements, i.e., answered ``completely'' or ``to a great extent'' (number on the left of the bars). The percentages do not consider the answers marked as ``I don't know.''}
    \label{fig:followup}
    \vspace{-10px}
\end{figure*}

Fourteen respondents evaluated a set of statements about how much our findings hold true in 2024. The results, presented in Fig.~\ref{fig:followup}, show that more than 75\% of the respondents believe that six out of the seven findings remain completely or largely relevant in 2024 (see numbers on the left of the bars). The finding with the least support refers to the majority groups perceiving D\&I initiatives as segregative instead of constructive.

We also invited participants to share their insights about feeling welcome in OSS. One respondent highlighted that stereotyping persists as \textit{``the implicit feeling that people aren't welcome if they don't look like 90\% of the maintainers.''} Another respondent suggested that future studies should analyze the \textit{``language barrier for non-native speakers''} as an interpersonal challenge affecting the feeling of welcomeness.

In summary, this follow-up study suggests that the findings from the analysis of the 2021 survey remain current at the time this paper was written.

%Table \ref{tab:persistent} shows the perception of OSS contributors regarding which of the interpersonal challenges investigated in 2021 still persist in 2024.
%\input{Tables/persisting_challenges}

% As discussed in Section \ref{sec:method_followup}), we conducted a follow-up study in 2024 to gather community perceptions of our result and the challenges identified in our dataset from 2021. This was especially important not only for triangulating our findings but also for assessing the persistence of these challenges over time.

%In the second part of the study, we conducted a poll during a keynote at a practitioner conference. In the first question, 56\% of the respondents considered the interpersonal challenges \textsc{significantly} influence the feeling welcome of contributors, followed by 28\% who considered the challenges to have extreme influence, 12\% considered a moderate influence and 4\% considered a slight influence. The second question showed that \textsc{lack of response}, \textsc{harassment}, and \textsc{stereotyping} were frequently mentioned.

% }

%Results indicate that, although a few years old, the dataset we used in the present study, which is the largest-scale survey on diversity and inclusion in OSS so far, still provides relevant data to explore in the context of challenges and feelings of being welcome in OSS. 

\vspace{-3px}
\section{Discussion}
\label{sec:discussion}
\vspace{-2px}

%\subsection{Discussion of Findings}

We investigated the relationship between \textit{interpersonal challenges} and the \textit{feeling of welcomeness} in OSS projects. 
%We looked at different challenges that are critical to contributors in the OSS domain. 
Interpersonal challenges are negatively associated with not feeling welcome, even more strongly for gender minorities and people with disabilities. Results align with previous research that discussed that contributors do not feel welcome when facing challenges \cite{steinmacher2015social,storey2016social}. In the following, we bringing exemplar quotes from the respondents' responses.

\textbf{Interactions and conflicts.} In RQ3, we found a correlation between \textsc{conflict or tension with other contributors} and \textsc{unwelcoming language} and between \textsc{conflict or tension with other contributors} and \textsc{stereotyping}, considering the three minority groups of gender, race, and disabilities. Those conflicts can happen through uncivil comments and toxic discussions in mailing lists, comments in pull-requests \cite{miller2022did}, or also during in-person events. 

\textbf{The delicate balance of welcoming minorities without appearing unwelcoming to majorities:}
Stereotyping is a fixed, oversimplified, and often biased belief about a group of people \cite{mccauley1980stereotyping}. Our results from RQ3 showed that gender minorities and people with disabilities have higher odds of experiencing stereotyping than their counterparts (Table \ref{tab:logistic}). However, in Sec. \ref{sec:results_rq4}, we found that stereotyping hurts regardless if they are part of a minority (e.g., ``I feel unwelcome because I'm a woman'') or majority (e.g., ``I'm a white guy. Everyone hates me.''). We noticed that minorities and majorities perceive that somehow \textit{``diversity is over pushed."} (P56). As noted by P218, \textit{``I never felt unwelcome in OSS until this push for inclusion and diversity. I am an Asian female mixed, Greek Yankee and Japanese. I never felt unwelcome in the '90s or early 2000s.''} Diversity, Equity, Inclusion, and accessibility initiatives must navigate carefully to avoid societal division. While focusing on demographic diversity, they often overlook the deeper cultural shifts required for genuine equity and inclusion. Mere increases in minority representation don't guarantee progress. True advancement demands transforming organizational culture. To gain public acceptance, we must reject dualistic thinking and emphasize universal inclusion \cite{friedman2024paradox}.

%Our results show some signs that \textbf{maintainers are overloaded}, burning out. As it is possible to notice in Figure~\ref{fig:freq_challenges}, there is a high incidence of people who had experienced \textsc{lack of response or rejection of contributions}. This resonates with previous literature showing that burnout is a thing in software development \cite{trinkenreich2023seip} and that, specifically in OSS, maintainers' responsibilities are too broad~\cite{dias2021makes,raman2020stress} resulting in stress and burnout~\cite{raman2020stress, geiger2021labor}. More research would be needed to track the deep roots of this phenomenon and understand the reasons why only approximately 1/3 of the respondents reported that they never faced this challenge. Although communities can monitor pull requests and mailing lists to guarantee that members' posts are not being missed \cite{miller2022did}, maintainers also need help with more maintainers to share their workload.

Our results show that \textbf{contributors miss feedback on their questions or acceptance of their contributions}. As Fig.~\ref{fig:freq_challenges} shows, there is a high incidence of people who had experienced \textsc{lack of response or rejection of contributions}. The reason for this phenomenon can be related to overloaded maintainers with several responsibilities~\cite{dias2021makes,raman2020stress}, as a result, they have a hard time answering contributors' requests. More research is required to track the deep roots of why only approximately $1/3$ of the respondents reported that they never faced this challenge. Although communities can monitor pull requests and mailing lists to guarantee that members' posts are not being missed \cite{miller2022did}, more maintainers are needed to share their workload. Generative AI can be instrumental in providing better feedback and answers to contributors \cite{10.1145/3664646.3664758,abedu2024llm}.

\textbf{The trifecta impeding inclusion: contribution-bias, identity hiding, and doxxing:} Diversity and inclusion initiatives are increasingly implemented in OSS projects~\cite{guizani2022perceptions}, but people's culture still needs to change so minorities are truly included, as mentioned by one of the respondents: \textit{``anti-women attitudes are rampant in popular open source projects, it is completely unacceptable, and yet it is tolerated"}. When knowing that discrimination is still tolerated, gender minorities often opt to hide their identities and create pseudonyms to avoid judgment~\cite{lee2019floss} to avoid discrimination and code review bias \cite{terrell2017gender,vasilescu2015perceptions,kofink2015contributions,canedo2020work}. This behavior was also observed by \citet{ford2019beyond} in online communities, where participants use a ``gender neutral alias for websites like technical communities, because [they] get better help when asking questions or answering them."
%As we show in Table \ref{tab:mixed}, when including the random effect of \textsc{doxxing}, the odds ratio of feeling welcome by the gender minorities in our sample is even more reduced. 
P1087 (men) claimed for \textit{``more anonymity, not judging people based on their irrelevant features."} However, one's identity should not be considered irrelevant. 
As we showed in Fig.~\ref{fig:freq_challenges}, \textsc{doxxing} was more frequently experienced by gender minorities than men in our sample. Although people may perceive one's identity as not relevant, when not revealing their identity, gender minorities may suffer an even more severe disadvantage in survival probability \cite{canedo2020work}. Although it prevents discrimination, hiding identity can lead to a lack of trust and ultimately cause a higher exit rate for such users \cite{trinkenreich2022women}. 

\textbf{The pervasive set of challenges still affecting gender minorties} \textsc{Sexual harassment}, \cite{singh2019women,nafus2012patches}, \textsc{stalking}, and \textsc{stereotyping} were the top three interpersonal challenges with higher odds to be experienced by gender minorities (see Table \ref{tab:logistic}. Although most communities have strict rules against negative behaviors, the enforcement of the code of conduct is crucial, as is the need for proactive measures to detect and prevent such challenges from adversely affecting contributors. The code of conduct comprises the collective norms of a community, as mantras that shape the culture of collaboration~ as well as the community’s expectations and values to create a friendly and welcoming community~\cite{tourani2017code}. Communities can use mining tools to identify gender pronouns in messages such as mailing lists, pull requests, and code reviews. Actions may include providing online training with practical examples of acceptable and non-acceptable behaviors and training for allies to speak up and act as ``collaborators, accomplices, and co-conspirators''~\cite{melaku2020better}. Additionally, lawyers should be involved in creating an enforcement plan and adding the appropriate terms to provide transparency about the punishments for those who violate the code of conduct.

\textbf{Implications to practice:} The insights from our research offer practical guidance for OSS communities striving to become more inclusive. For instance, our findings on the particular challenges faced by gender minorities can inform the development of targeted mentorship programs, codes of conduct, and communication guidelines that specifically address these issues. Similarly, the unique experiences of racial minorities and people with disabilities in OSS highlighted in our study could lead to the implementation of community guidelines that promote respect and understanding across diverse backgrounds. By focusing on reducing interpersonal challenges, it is possible that the contributors feel more welcome, and, ultimately, OSS projects can enhance their attractiveness to a broader range of contributors, enriching the community and the software products they develop.

Our findings can inform communities in prioritizing interventions such as training, mentorship, and support programs. For example, since interpersonal challenges were more strongly associated with the lack of feeling welcome for gender minorities and people with disabilities (RQ2), communities aiming to improve the feelings of welcomeness of gender minorities should combat interpersonal challenges, especially sexual harassment, stalking, and doxxing, which our data (RQ3) showed are currently prevalent among minorities. Potential actions can be to have an enforceable Code of Conduct, promote allies who can voice concerns without direct retribution, and set up Diversity, Equity, and Inclusion committees.

\textbf{Implications to research:} While our study sheds light on critical aspects of diversity in OSS, it also opens avenues for further research. Future research could explore longitudinal trends in the experiences of minority groups within OSS communities, providing insights into how inclusivity efforts evolve and their long-term effectiveness. Examining the intersectionality of various demographic factors could offer a more comprehensive understanding of the challenges faced by individuals belonging to multiple underrepresented groups. Investigating how different strategies already in place (e.g., code of conduct, near-peer mentoring, demographic-specific support groups) impact the perception of welcomeness by conducting case studies and in-depth qualitative investigations. Another interesting future work would assess the association of welcomeness and sense of belonging---given their theoretical relationship~\cite{hailey2022racialized}.

\section{Threats to Validity}
\label{sec:threats}
\vspace{-2px}

%This study has a number of limitations, which we discuss next. 

We analyzed the answers to an existing survey conducted by the Linux Foundation Research. The survey was more extensive than what was analyzed in the context of this paper. Another option would be conducting another survey. We opted to use the existing dataset given the breadth and the number of answers of this one, which would not be possible since recruitment in software engineering has been an issue recently~\cite{alami2024you, danilova2021you}.
The constructs under analysis are captured by the survey, and one of the authors took part in the design of the instrument, which was tailored to existing measurement instruments for the constructs based on prior literature. Our analysis confirmed that the adopted constructs were internally consistent and scored satisfactorily on convergent and discriminant validity tests.

The hypothesis tested in this study proposes an association between different constructs rather than causal relationships, as the present study is a cross-sectional sample study \cite{stol2018abc}. We acknowledge the limitation that our respondents comprise contributors who are more likely to feel welcome as they dedicated their time to answering the questionnaire, suggesting a response bias. While it is clear that contributors motivated by some intrinsic-social reasons tend to feel more welcome, a theoretical model such as ours cannot capture a complete and exhaustive list of factors. Moreover, feeling welcome was measured on a single question, and the nuanced complexity of this feeling could benefit by having a latent construct variable with multiple questions. Other factors can play a role, and our results represent a starting point for future studies.

The respondent demographics align with the overall distribution of previous measurements of OSS contributors, making this a suitable starting point for understanding the link between interpersonal challenges and feeling welcome. Further replication studies in individual communities can replicate, validate, and extend our theoretical model. According to previous research, most OSS contributors are men, white, and without disabilities, consistent with our sample. The responses were sufficiently consistent to find full support for the hypothesis. 

The survey was conducted in 2021 and provides a rich and comprehensive perspective rarely seen in academic studies. However, since 2021, there may have been changes in the challenges OSS contributors face and their perceptions of inclusivity. To address this threat, we conducted a follow-up survey with practitioners (see Sec. \ref{sec:results_followup}). 
Eighty-nine percent of respondents confirmed that the challenges we investigated still substantially influence the sense of welcome in OSS as of 2024, when this paper was written. None said these challenges do not influence welcomeness.
Future longitudinal research is needed to investigate how our results may change over time as OSS culture slowly changes.

\vspace{-4pt}
\section{Conclusion}
\label{sec:conclusion}
%\vspace{-3px}

We assessed the interplay between interpersonal challenges and the perception of welcomeness in OSS communities. Our findings indicate a significant negative correlation between interpersonal challenges and feelings of welcomeness, with pronounced effects on gender minorities and people with disabilities. While these results might not surprise practitioners, our study reveals the strengths of the associations between each challenge and different demographic groups. These findings hold significant implications for communities aiming to enhance welcomeness among their members by focusing on the challenges faced by diverse groups. This involves more than just implementing inclusive policies; it requires cultivating a culture that actively opposes bias, champions diversity, and fosters a sense of belonging and respect. 
%This aligns with the core ethos of openness and collaboration in OSS and enhances the richness and innovation within these communities by embracing more perspectives and experiences. 

\vspace{-4pt}
\section{Acknowledgements}
We want to thank the respondents and The Linux Foundation Research for making the survey dataset available. This work was partially supported by the National Science Foundation grants 2235601, 2236198, 2247929, and 2303042. Any opinions expressed in this material are those of the authors and do not necessarily reflect the views of The Linux Foundation.

\bibliographystyle{IEEEtranN}
\footnotesize{\bibliography{REFERENCES}}

% Generated by IEEEtranN.bst, version: 1.14 (2015/08/26)
\begin{thebibliography}{91}
\providecommand{\natexlab}[1]{#1}
\providecommand{\url}[1]{#1}
\csname url@samestyle\endcsname
\providecommand{\newblock}{\relax}
\providecommand{\bibinfo}[2]{#2}
\providecommand{\BIBentrySTDinterwordspacing}{\spaceskip=0pt\relax}
\providecommand{\BIBentryALTinterwordstretchfactor}{4}
\providecommand{\BIBentryALTinterwordspacing}{\spaceskip=\fontdimen2\font plus
\BIBentryALTinterwordstretchfactor\fontdimen3\font minus
  \fontdimen4\font\relax}
\providecommand{\BIBforeignlanguage}[2]{{%
\expandafter\ifx\csname l@#1\endcsname\relax
\typeout{** WARNING: IEEEtranN.bst: No hyphenation pattern has been}%
\typeout{** loaded for the language `#1'. Using the pattern for}%
\typeout{** the default language instead.}%
\else
\language=\csname l@#1\endcsname
\fi
#2}}
\providecommand{\BIBdecl}{\relax}
\BIBdecl

\bibitem[Trinkenreich et~al.(2022)Trinkenreich, Wiese, Sarma, Gerosa, and
  Steinmacher]{trinkenreich2022women}
B.~Trinkenreich, I.~Wiese, A.~Sarma, M.~Gerosa, and I.~Steinmacher, ``Women’s
  participation in open source software: A survey of the literature,''
  \emph{ACM Trans Soft Eng Methodol}, vol.~31, no.~4, 2022.

\bibitem[Nadri et~al.(2021{\natexlab{a}})Nadri, Rodr{\'\i}guez-P{\'e}rez, and
  Nagappan]{nadri2021relationship}
R.~Nadri, G.~Rodr{\'\i}guez-P{\'e}rez, and M.~Nagappan, ``On the relationship
  between the developer’s perceptible race and ethnicity and the evaluation
  of contributions in oss,'' \emph{IEEE Transactions on Software Engineering},
  vol.~48, no.~8, pp. 2955--2968, 2021.

\bibitem[Carter and Groopman(2021)]{lfdiversitysurvey}
H.~Carter and J.~Groopman, ``The linux foundation report on diversity, equity,
  and inclusion in open source,'' 2021, [accessed 2023-06-27].

\bibitem[Vasilescu et~al.(2015{\natexlab{a}})Vasilescu, Posnett, Ray, van~den
  Brand, Serebrenik, Devanbu, and Filkov]{vasilescu2015gender}
B.~Vasilescu, D.~Posnett, B.~Ray, M.~G. van~den Brand, A.~Serebrenik,
  P.~Devanbu, and V.~Filkov, ``Gender and tenure diversity in github teams,''
  in \emph{Proceedings of the 33rd annual ACM conference on human factors in
  computing systems}, 2015, pp. 3789--3798.

\bibitem[Bosu and Sultana(2019)]{bosu2019diversity}
A.~Bosu and K.~Z. Sultana, ``Diversity and inclusion in open source software
  (oss) projects: Where do we stand?'' in \emph{2019 ACM/IEEE International
  Symposium on Empirical Software Engineering and Measurement (ESEM)}.\hskip
  1em plus 0.5em minus 0.4em\relax IEEE, 2019, pp. 1--11.

\bibitem[Steinmacher et~al.(2015{\natexlab{a}})Steinmacher, Conte, Gerosa, and
  Redmiles]{steinmacher2015social}
I.~Steinmacher, T.~Conte, M.~A. Gerosa, and D.~Redmiles, ``Social barriers
  faced by newcomers placing their first contribution in open source software
  projects,'' in \emph{18th ACM conference on Computer supported cooperative
  work \& social computing}, 2015, pp. 1379--1392.

\bibitem[Steinmacher et~al.(2015{\natexlab{b}})Steinmacher, Conte, and
  Gerosa]{Steinmacher.Conte.ea_2015HICSS}
I.~Steinmacher, T.~Conte, and M.~A. Gerosa, ``Understanding and supporting the
  choice of an appropriate task to start with in open source software
  communities,'' in \emph{HICSS'15}.\hskip 1em plus 0.5em minus 0.4em\relax
  IEEE, 2015, pp. 5299--5308.

\bibitem[Jensen et~al.(2011)Jensen, King, and Kuechler]{jensen2011joining}
C.~Jensen, S.~King, and V.~Kuechler, ``Joining free/open source software
  communities: An analysis of newbies' first interactions on project mailing
  lists,'' in \emph{HICSS 2011}.\hskip 1em plus 0.5em minus 0.4em\relax IEEE,
  2011, pp. 1--10.

\bibitem[Hannebauer and Gruhn(2017)]{hannebauer2017relationship}
C.~Hannebauer and V.~Gruhn, ``On the relationship between newcomer motivations
  and contribution barriers in open source projects,'' in \emph{International
  Symposium on Open Collaboration}, 2017, pp. 1--10.

\bibitem[Pinto et~al.(2016)Pinto, Steinmacher, and Gerosa]{pinto2016more}
G.~Pinto, I.~Steinmacher, and M.~A. Gerosa, ``More common than you think: An
  in-depth study of casual contributors,'' in \emph{2016 IEEE 23rd
  International Conference on Software Analysis, Evolution, and Reengineering
  (SANER)}, vol.~1.\hskip 1em plus 0.5em minus 0.4em\relax IEEE, 2016, pp.
  112--123.

\bibitem[Balali et~al.(2018)Balali, Steinmacher, Annamalai, Sarma, and
  Gerosa]{balali2018mentor}
S.~Balali, I.~Steinmacher, U.~Annamalai, A.~Sarma, and M.~A. Gerosa,
  ``Newcomers' barriers... is that all? an analysis of mentors' and newcomers'
  barriers in {OSS} projects,'' \emph{Computer Supported Cooperative Work},
  vol.~27, no. 3-6, pp. 679--714, 2018.

\bibitem[Trinkenreich et~al.(2023)Trinkenreich, Stol, Sarma, German, Gerosa,
  and Steinmacher]{trinkenreich2023belong}
B.~Trinkenreich, K.-J. Stol, A.~Sarma, D.~German, M.~Gerosa, and
  I.~Steinmacher, ``Do i belong? modeling sense of virtual community among
  linux kernel contributors,'' in \emph{ICSE 2023}.\hskip 1em plus 0.5em minus
  0.4em\relax IEEE, 2023.

\bibitem[Storey et~al.(2016)Storey, Zagalsky, Figueira~Filho, Singer, and
  German]{storey2016social}
M.-A. Storey, A.~Zagalsky, F.~Figueira~Filho, L.~Singer, and D.~M. German,
  ``How social and communication channels shape and challenge a participatory
  culture in software development,'' \emph{IEEE Transactions on Software
  Engineering}, vol.~43, no.~2, pp. 185--204, 2016.

\bibitem[Rodr{\'\i}guez-P{\'e}rez et~al.(2021)Rodr{\'\i}guez-P{\'e}rez, Nadri,
  and Nagappan]{rodriguez2021perceived}
G.~Rodr{\'\i}guez-P{\'e}rez, R.~Nadri, and M.~Nagappan, ``Perceived diversity
  in software engineering: a systematic literature review,'' \emph{Empirical
  Software Engineering}, vol.~26, pp. 1--38, 2021.

\bibitem[Nadri et~al.(2021{\natexlab{b}})Nadri, Rodriguez-Perez, and
  Nagappan]{nadri2021insights}
R.~Nadri, G.~Rodriguez-Perez, and M.~Nagappan, ``Insights into nonmerged pull
  requests in github: Is there evidence of bias based on perceptible race?''
  \emph{IEEE Softw.}, vol.~38, no.~2, pp. 51--57, 2021.

\bibitem[Forte and Lampe(2013)]{forte2013defining}
A.~Forte and C.~Lampe, ``Defining, understanding, and supporting open
  collaboration: Lessons from the literature,'' \emph{American Behavioral
  Scientist}, vol.~57, no.~5, pp. 535--547, 2013.

\bibitem[Lee and Carver(2019)]{lee2019floss}
A.~Lee and J.~C. Carver, ``Floss participants' perceptions about gender and
  inclusiveness: a survey,'' in \emph{ICSE 2019}.\hskip 1em plus 0.5em minus
  0.4em\relax IEEE, 2019, pp. 677--687.

\bibitem[{The Linux Foundation Research}(2021{\natexlab{a}})]{dei_survey}
{The Linux Foundation Research}, ``Diversity, equity, and inclusion in open
  source,''
  \url{https://www.linuxfoundation.org/research/the-2021-linux-foundation-report-on-diversity-equity-and-inclusion-in-open-source?hsLang=en},
  2021, [Online; accessed 2023-07-19].

\bibitem[Creswell and Creswell(2017)]{creswell2017research}
J.~W. Creswell and J.~D. Creswell, \emph{Research design: Qualitative,
  quantitative, and mixed methods approaches}.\hskip 1em plus 0.5em minus
  0.4em\relax Sage publications, 2017.

\bibitem[Feng(2023)]{feng2023state}
Z.~Feng, ``The state of survival in oss: The impact of diversity,'' in
  \emph{FSE 2023}, 2023, pp. 2213--2215.

\bibitem[Feng et~al.(2023)Feng, Guizani, Gerosa, and Sarma]{feng2023state1}
Z.~Feng, M.~Guizani, M.~A. Gerosa, and A.~Sarma, ``The state of diversity and
  inclusion in {A}pache: A pulse check,'' in \emph{CHASE 2023}.\hskip 1em plus
  0.5em minus 0.4em\relax IEEE, 2023, pp. 150--160.

\bibitem[Trinkenreich et~al.(2020)Trinkenreich, Guizani, Wiese, Sarma, and
  Steinmacher]{trinkenreich2020hidden}
B.~Trinkenreich, M.~Guizani, I.~Wiese, A.~Sarma, and I.~Steinmacher, ``Hidden
  figures: Roles and pathways of successful oss contributors,''
  \emph{Proceedings of the ACM on Human-Computer Interaction}, vol.~4, no.
  CSCW2, pp. 1--22, 2020.

\bibitem[Trinkenreich et~al.(2021)Trinkenreich, Guizani, Wiese, Conte, Gerosa,
  Sarma, and Steinmacher]{trinkenreich2021pot}
B.~Trinkenreich, M.~Guizani, I.~Wiese, T.~Conte, M.~Gerosa, A.~Sarma, and
  I.~Steinmacher, ``Pots of gold at the end of the rainbow: What is success for
  open source contributors?'' \emph{IEEE Transactions on Software Engineering},
  vol.~48, no.~10, pp. 3940--3953, 2021.

\bibitem[Mendez et~al.(2018)Mendez, Sarma, and Burnett]{Mendez2018GE}
\BIBentryALTinterwordspacing
C.~J. Mendez, A.~Sarma, and M.~Burnett, ``Gender in open source software: What
  the tools tell,'' in \emph{2018 {IEEE/ACM} 1st International Workshop on
  Gender Equality in Software Engineering, GE@ICSE, Gothenburg, Sweden, May 28,
  2018}, 2018, pp. 21--24. [Online]. Available:
  \url{http://ieeexplore.ieee.org/document/8452746}
\BIBentrySTDinterwordspacing

\bibitem[Guizani et~al.(2022)Guizani, Trinkenreich, Castro-Guzman, Steinmacher,
  Gerosa, and Sarma]{guizani2022perceptions}
M.~Guizani, B.~Trinkenreich, A.~A. Castro-Guzman, I.~Steinmacher, M.~Gerosa,
  and A.~Sarma, ``Perceptions of the state of d\&i and d\&i initiative in the
  asf,'' in \emph{ICSE-SEIS 2022}, 2022, pp. 130--142.

\bibitem[Huang et~al.(2021)Huang, Ford, and Zimmermann]{huang2021leaving}
Y.~Huang, D.~Ford, and T.~Zimmermann, ``Leaving my fingerprints: Motivations
  and challenges of contributing to oss for social good,'' in \emph{ICSE
  2021}.\hskip 1em plus 0.5em minus 0.4em\relax IEEE, 2021, pp. 1020--1032.

\bibitem[{Apache Software Foundation}(2023)]{apache2023dei}
\BIBentryALTinterwordspacing
{Apache Software Foundation}, ``2023 diversity, equity, and inclusion report,''
  2023, accessed: 2024-03-14. [Online]. Available:
  \url{https://www.apache.org/foundation/docs/2023DEIReport.pdf}
\BIBentrySTDinterwordspacing

\bibitem[Finley(2023)]{DiversityOSS2023}
K.~Finley, ``Diversity in open source is even worse than in tech overall,''
  \url{https://www.wired.com/2017/06/diversity-open-source-even-worse-tech-overall/},
  2023, accessed: 2024-03-19.

\bibitem[Miller et~al.(2022)Miller, Cohen, Klug, Vasilescu, and
  KaUstner]{miller2022did}
C.~Miller, S.~Cohen, D.~Klug, B.~Vasilescu, and C.~KaUstner, ``" did you miss
  my comment or what?" understanding toxicity in open source discussions,'' in
  \emph{ICSE 2022}, 2022, pp. 710--722.

\bibitem[Ferreira et~al.(2021)Ferreira, Cheng, and Adams]{ferreira2021shut}
I.~Ferreira, J.~Cheng, and B.~Adams, ``The" shut the f** k up" phenomenon:
  Characterizing incivility in open source code review discussions,''
  \emph{Proceedings of the ACM on Human-Computer Interaction}, vol.~5, no.
  CSCW2, pp. 1--35, 2021.

\bibitem[Powell et~al.(2010)Powell, Hunsinger, and Medlin]{powell2010gender}
W.~E. Powell, D.~S. Hunsinger, and B.~D. Medlin, ``Gender differences within
  the open source community: An exploratory study,'' \emph{Journal of
  Information Technology}, vol.~21, no.~4, pp. 29--37, 2010.

\bibitem[Vasilescu et~al.(2015{\natexlab{b}})Vasilescu, Filkov, and
  Serebrenik]{vasilescu2015perceptions}
B.~Vasilescu, V.~Filkov, and A.~Serebrenik, ``Perceptions of diversity on git
  hub: A user survey,'' in \emph{CHASE 2015}.\hskip 1em plus 0.5em minus
  0.4em\relax IEEE, 2015, pp. 50--56.

\bibitem[Singh and Brandon(2022)]{singh2022discrimination}
V.~Singh and W.~Brandon, ``Discrimination, misogyny and harassment: Examples
  from oss: content analysis of women-focused online discussion forums,'' in
  \emph{Proceedings of the Third Workshop on Gender Equality, Diversity, and
  Inclusion in Software Engineering}, 2022, pp. 71--79.

\bibitem[Singh et~al.(2021)Singh, Bongiovanni, and Brandon]{singh2021codes}
V.~Singh, B.~Bongiovanni, and W.~Brandon, ``Codes of conduct in open source
  software—for warm and fuzzy feelings or equality in community?''
  \emph{Software Quality Journal}, pp. 1--40, 2021.

\bibitem[Hailey(2022)]{hailey2022racialized}
C.~A. Hailey, ``Racialized perceptions of anticipated school belonging,''
  \emph{Educational Policy}, vol.~36, no.~4, pp. 879--910, 2022.

\bibitem[Gill(2018)]{gill2018suppression}
N.~Gill, ``The suppression of welcome,'' \emph{Fennia-International Journal of
  Geography}, vol. 196, no.~1, pp. 88--98, 2018.

\bibitem[Foor et~al.(2007)Foor, Walden, and Trytten]{foor2007wish}
C.~E. Foor, S.~E. Walden, and D.~A. Trytten, ``“i wish that i belonged more
  in this whole engineering group:” achieving individual diversity,''
  \emph{Journal of Engineering Education}, vol.~96, no.~2, pp. 103--115, 2007.

\bibitem[Bopp et~al.(2017)Bopp, Turick, Vadeboncoeur, and Aicher]{bopp2017you}
T.~Bopp, R.~Turick, J.~D. Vadeboncoeur, and T.~J. Aicher, ``Are you welcomed? a
  racial and ethnic comparison of perceived welcomeness in sport
  participation,'' \emph{International Journal of Exercise Science}, vol.~10,
  no.~6, pp. 833--844, 2017.

\bibitem[Pitterson et~al.(2022)Pitterson, Streveler, and
  Douglas]{pitterson2022helping}
N.~Pitterson, R.~Streveler, and T.~C. Douglas, ``Helping newcomers feel “at
  home”: Reflections on the start of careers in engineering education
  research and implications for broadening participation,'' in \emph{2022 IEEE
  Frontiers in Education Conference (FIE)}.\hskip 1em plus 0.5em minus
  0.4em\relax IEEE, 2022, pp. 1--6.

\bibitem[Williams(2015)]{williams2015beyond}
L.~M. Williams, ``Beyond enforcement: Welcomeness, local law enforcement, and
  immigrants,'' \emph{Public Administration Review}, vol.~75, no.~3, pp.
  433--442, 2015.

\bibitem[Wen et~al.(2007)Wen, Hudak, and Hwang]{wen2007homeless}
C.~K. Wen, P.~L. Hudak, and S.~W. Hwang, ``Homeless people’s perceptions of
  welcomeness and unwelcomeness in healthcare encounters,'' \emph{Journal of
  General Internal Medicine}, vol.~22, pp. 1011--1017, 2007.

\bibitem[Gren and Ralph(2022)]{gren2022makes}
L.~Gren and P.~Ralph, ``What makes effective leadership in agile software
  development teams?'' in \emph{ICSE 2022}, 2022, pp. 2402--2414.

\bibitem[Trinkenreich et~al.(2024)Trinkenreich, Gerosa, and
  Steinmacher]{trinkenreich2024unraveling}
B.~Trinkenreich, M.~A. Gerosa, and I.~Steinmacher, ``Unraveling the drivers of
  sense of belonging in software delivery teams: Insights from a large-scale
  survey,'' in \emph{ICSE 2024}.\hskip 1em plus 0.5em minus 0.4em\relax IEEE,
  2024, pp. 974--974.

\bibitem[{The Linux Foundation Research}(2021{\natexlab{b}})]{dei_survey_data}
{The Linux Foundation Research}, ``Survey data --- 2021 diversity, equity, and
  inclusion in open source,''
  \url{https://data.world/thelinuxfoundation/2021-diversity-equity-and-inclusion-in-open-source},
  2021, [Online; accessed 2023-07-19].

\bibitem[Hair et~al.(2017)Hair, Hollingsworth, Randolph, and
  Chong]{hair2017updated}
J.~Hair, C.~L. Hollingsworth, A.~B. Randolph, and A.~Y.~L. Chong, ``An updated
  and expanded assessment of pls-sem in information systems research,''
  \emph{Industrial management \& data systems}, 2017.

\bibitem[Blind(2023)]{replication_package}
Blind, ``Replication package,''
  \url{https://figshare.com/s/9b0763fac157e95493be}, 2023.

\bibitem[Bembich and Bortolotti(2017)]{bembich2017early}
C.~Bembich and E.~Bortolotti, ``Early childhood services: Improving the
  relational and emotional skills of educators to ensure inclusiveness,''
  \emph{Early Childhood Services: improving the relational and emotional skills
  of educators to ensure inclusiveness}, pp. 30--44, 2017.

\bibitem[Stewart et~al.(2019)Stewart, Ju, and Kendrick]{stewart2019racial}
B.~Stewart, B.~Ju, and K.~D. Kendrick, ``Racial climate and inclusiveness in
  academic libraries: perceptions of welcomeness among black college
  students,'' \emph{The Library Quarterly}, vol.~89, no.~1, pp. 16--33, 2019.

\bibitem[Foundation(2021)]{lfcodeofconduct}
T.~L. Foundation, ``The linux foundation code of conduct,'' 2021, [online;
  accessed 2024-07-02].

\bibitem[Ringle et~al.(2015)Ringle, Da~Silva, and Bido]{ringle2015structural}
C.~Ringle, D.~Da~Silva, and D.~Bido, ``Structural equation modeling with the
  smartpls,'' \emph{Bido, D., da Silva, D., \& Ringle, C.(2014). Structural
  Equation Modeling with the Smartpls. Brazilian Journal Of Marketing},
  vol.~13, no.~2, 2015.

\bibitem[Dubberley et~al.(2020)Dubberley, Koenig, and
  Murray]{dubberley2020digital}
S.~Dubberley, A.~Koenig, and D.~Murray, \emph{Digital witness: using open
  source information for human rights investigation, documentation, and
  accountability}.\hskip 1em plus 0.5em minus 0.4em\relax Oxford University
  Press, USA, 2020.

\bibitem[Nafus(2012)]{nafus2012patches}
D.~Nafus, ``‘patches don’t have gender’: What is not open in open source
  software,'' \emph{New Media \& Society}, vol.~14, no.~4, pp. 669--683, 2012.

\bibitem[Pittaro(2007)]{pittaro2007cyber}
M.~L. Pittaro, ``Cyber stalking: An analysis of online harassment and
  intimidation,'' \emph{International journal of cyber criminology}, vol.~1,
  no.~2, pp. 180--197, 2007.

\bibitem[Brady and Nobles(2017)]{brady2017dark}
P.~Q. Brady and M.~R. Nobles, ``The dark figure of stalking: Examining law
  enforcement response,'' \emph{Journal of interpersonal violence}, vol.~32,
  no.~20, pp. 3149--3173, 2017.

\bibitem[Tourani et~al.(2017)Tourani, Adams, and Serebrenik]{tourani2017code}
P.~Tourani, B.~Adams, and A.~Serebrenik, ``Code of conduct in open source
  projects,'' in \emph{2017 IEEE 24th international conference on software
  analysis, evolution and reengineering (SANER)}.\hskip 1em plus 0.5em minus
  0.4em\relax IEEE, 2017, pp. 24--33.

\bibitem[McCauley et~al.(1980)McCauley, Stitt, and
  Segal]{mccauley1980stereotyping}
C.~McCauley, C.~L. Stitt, and M.~Segal, ``Stereotyping: From prejudice to
  prediction.'' \emph{Psychological Bulletin}, vol.~87, no.~1, p. 195, 1980.

\bibitem[Paul et~al.(2019)Paul, Bosu, and Sultana]{paul2019expressions}
R.~Paul, A.~Bosu, and K.~Z. Sultana, ``Expressions of sentiments during code
  reviews: Male vs. female,'' in \emph{2019 IEEE 26th International Conference
  on Software Analysis, Evolution and Reengineering (SANER)}.\hskip 1em plus
  0.5em minus 0.4em\relax IEEE, 2019, pp. 26--37.

\bibitem[Qiu et~al.(2019)Qiu, Li, Padala, Sarma, and Vasilescu]{qiu2019signals}
H.~S. Qiu, Y.~L. Li, S.~Padala, A.~Sarma, and B.~Vasilescu, ``The signals that
  potential contributors look for when choosing open-source projects,''
  \emph{Proceedings of the ACM on Human-Computer Interaction}, vol.~3, pp.
  1--29, 2019.

\bibitem[Robles et~al.(2014)Robles, Arjona, Serebrenik, Vasilescu, and
  Gonz{\'a}lez-Barahona]{robles2014floss}
G.~Robles, L.~Arjona, A.~Serebrenik, B.~Vasilescu, and
  J.~Gonz{\'a}lez-Barahona, ``Floss 2013: A survey dataset about free software
  contributors: challenges for curating, sharing, and combining,'' in
  \emph{Conference on Mining Software Repositories}, 2014, pp. 396--399.

\bibitem[Guizani et~al.(2021)Guizani, Chatterjee, Trinkenreich, May, Noa,
  Russell, Zambrano, Izquierdo-Cortazar, Steimacher, Gerosa, and
  Sarma]{guizani2021}
M.~Guizani, A.~Chatterjee, B.~Trinkenreich, M.~May, G.~Noa, L.~Russell,
  G.~Zambrano, D.~Izquierdo-Cortazar, I.~Steimacher, M.~Gerosa, and A.~Sarma,
  ``The long road ahead: Ongoing challenges in contributing to large oss
  organizations and what to do,'' \emph{Computer Supported Cooperative Work},
  2021.

\bibitem[Kaur and Chahal(2022)]{kaur2022exploring}
R.~Kaur and K.~K. Chahal, ``Exploring factors affecting developer abandonment
  of open source software projects,'' \emph{Journal of Software: Evolution and
  Process}, vol.~34, no.~9, p. e2484, 2022.

\bibitem[Steinmacher et~al.(2013)Steinmacher, Wiese, Chaves, and
  Gerosa]{steinmacher2013newcomers}
I.~Steinmacher, I.~Wiese, A.~P. Chaves, and M.~A. Gerosa, ``Why do newcomers
  abandon open source software projects?'' in \emph{Cooperative and Human
  Aspects of Software Engineering (CHASE), 2013 6th International Workshop
  on}.\hskip 1em plus 0.5em minus 0.4em\relax IEEE, 2013, pp. 25--32.

\bibitem[Hair et~al.(2019)Hair, Risher, Sarstedt, and Ringle]{hair2019use}
J.~F. Hair, J.~J. Risher, M.~Sarstedt, and C.~M. Ringle, ``When to use and how
  to report the results of pls-sem,'' \emph{European business review}, 2019.

\bibitem[Albusays et~al.(2021)Albusays, Bjorn, Dabbish, Ford, Murphy-Hill,
  Serebrenik, and Storey]{albusays2021diversity}
K.~Albusays, P.~Bjorn, L.~Dabbish, D.~Ford, E.~Murphy-Hill, A.~Serebrenik, and
  M.-A. Storey, ``The diversity crisis in software development,'' \emph{IEEE
  Software}, vol.~38, no.~2, pp. 19--25, 2021.

\bibitem[Silveira and Prikladnicki(2019)]{silveira2019systematic}
K.~K. Silveira and R.~Prikladnicki, ``A systematic mapping study of diversity
  in software engineering: a perspective from the agile methodologies,'' in
  \emph{Cooperative and Human Aspects of Software Engineering}.\hskip 1em plus
  0.5em minus 0.4em\relax IEEE, 2019, pp. 7--10.

\bibitem[Hair~Jr et~al.(2017)Hair~Jr, Sarstedt, Ringle, and
  Gudergan]{hair2017advanced}
J.~F. Hair~Jr, M.~Sarstedt, C.~M. Ringle, and S.~P. Gudergan, \emph{Advanced
  issues in partial least squares structural equation modeling}.\hskip 1em plus
  0.5em minus 0.4em\relax saGe publications, 2017.

\bibitem[Henseler et~al.(2016{\natexlab{a}})Henseler, Ringle, and
  Sarstedt]{henseler2016testing}
J.~Henseler, C.~M. Ringle, and M.~Sarstedt, ``Testing measurement invariance of
  composites using partial least squares,'' \emph{International marketing
  review}, 2016.

\bibitem[Harrell and Harrell(2015)]{harrell2015ordinal}
F.~E. Harrell, Jr and F.~E. Harrell, ``Ordinal logistic regression,''
  \emph{Regression modeling strategies: with applications to linear models,
  logistic and ordinal regression, and survival analysis}, pp. 311--325, 2015.

\bibitem[SciPy(2024)]{SciPy2024}
SciPy, ``Scipy 2024 conference,'' \url{https://www.scipy2024.scipy.org/}, 2024,
  accessed: 2024-07-22.

\bibitem[Russo and Stol(2021)]{russo2021pls}
D.~Russo and K.-J. Stol, ``Pls-sem for software engineering research: An
  introduction and survey,'' \emph{ACM Computing Surveys (CSUR)}, vol.~54,
  no.~4, pp. 1--38, 2021.

\bibitem[Sarstedt and Cheah(2019)]{sarstedt2019partial}
M.~Sarstedt and J.-H. Cheah, ``Partial least squares structural equation
  modeling using smart{PLS}: a software review,'' 2019.

\bibitem[Henseler et~al.(2015)Henseler, Ringle, and Sarstedt]{henseler2015new}
J.~Henseler, C.~Ringle, and M.~Sarstedt, ``A new criterion for assessing
  discriminant validity in variance-based structural equation modeling,''
  \emph{Journal of the academy of marketing science}, vol.~43, no.~1, pp.
  115--135, 2015.

\bibitem[Henseler et~al.(2016{\natexlab{b}})Henseler, Hubona, and
  Ray]{henseler2016using}
J.~Henseler, G.~Hubona, and P.~Ray, ``Using pls path modeling in new technology
  research: updated guidelines,'' \emph{Ind. Manag. Data Syst.}, 2016.

\bibitem[Henseler et~al.(2014)Henseler, Dijkstra, Sarstedt, Ringle,
  Diamantopoulos, Straub, Ketchen~Jr, Hair, Hult, and
  Calantone]{henseler2014common}
J.~Henseler, T.~K. Dijkstra, M.~Sarstedt, C.~M. Ringle, A.~Diamantopoulos,
  D.~W. Straub, D.~J. Ketchen~Jr, J.~F. Hair, G.~T.~M. Hult, and R.~J.
  Calantone, ``Common beliefs and reality about pls: Comments on r{\"o}nkk{\"o}
  and evermann,'' \emph{Organ Res Methods}, vol.~17, no.~2, 2014.

\bibitem[Ringle and Sarstedt(2016)]{ringle2016gain}
C.~M. Ringle and M.~Sarstedt, ``Gain more insight from your pls-sem results:
  The importance-performance map analysis,'' \emph{Industrial management \&
  data systems}, 2016.

\bibitem[Thissen et~al.(2002)Thissen, Steinberg, and Kuang]{thissen2002quick}
D.~Thissen, L.~Steinberg, and D.~Kuang, ``Quick and easy implementation of the
  benjamini-hochberg procedure for controlling the false positive rate in
  multiple comparisons,'' \emph{Journal of educational and behavioral
  statistics}, vol.~27, no.~1, pp. 77--83, 2002.

\bibitem[Sedgwick(2014)]{sedgwick2014spearman}
P.~Sedgwick, ``Spearman’s rank correlation coefficient,'' \emph{Bmj}, vol.
  349, 2014.

\bibitem[Friedman and Vlady(2024)]{friedman2024paradox}
H.~H. Friedman and S.~Vlady, ``The paradox of dei: How lofty ideals became
  hated,'' \emph{North East Journal of Legal Studies}, vol.~44, no.~1, p.~4,
  2024.

\bibitem[Dias et~al.(2021)Dias, Meirelles, Castor, Steinmacher, Wiese, and
  Pinto]{dias2021makes}
E.~Dias, P.~Meirelles, F.~Castor, I.~Steinmacher, I.~Wiese, and G.~Pinto,
  ``What makes a great maintainer of open source projects?'' in \emph{ICSE
  2021}.\hskip 1em plus 0.5em minus 0.4em\relax IEEE, 2021, pp. 982--994.

\bibitem[Raman et~al.(2020)Raman, Cao, Tsvetkov, K{\"a}stner, and
  Vasilescu]{raman2020stress}
N.~Raman, M.~Cao, Y.~Tsvetkov, C.~K{\"a}stner, and B.~Vasilescu, ``Stress and
  burnout in open source: Toward finding, understanding, and mitigating
  unhealthy interactions,'' in \emph{42nd International Conference on Software
  Engineering: New Ideas and Emerging Results}, 2020, pp. 57--60.

\bibitem[Correia et~al.(2024)Correia, Nicholson, Coutinho, Barbosa,
  Castelluccio, Gerosa, Garcia, and Steinmacher]{10.1145/3664646.3664758}
\BIBentryALTinterwordspacing
J.~a. Correia, M.~C. Nicholson, D.~Coutinho, C.~Barbosa, M.~Castelluccio,
  M.~Gerosa, A.~Garcia, and I.~Steinmacher, ``Unveiling the potential of a
  conversational agent in developer support: Insights from {Mozilla’s}
  {PDF.js} project,'' in \emph{Proceedings of the 1st ACM International
  Conference on AI-Powered Software}, ser. AIware 2024.\hskip 1em plus 0.5em
  minus 0.4em\relax New York, NY, USA: Association for Computing Machinery,
  2024, p. 10–18. [Online]. Available:
  \url{https://doi.org/10.1145/3664646.3664758}
\BIBentrySTDinterwordspacing

\bibitem[Abedu et~al.(2024)Abedu, Abdellatif, and Shihab]{abedu2024llm}
S.~Abedu, A.~Abdellatif, and E.~Shihab, ``Llm-based chatbots for mining
  software repositories: Challenges and opportunities,'' in \emph{Proceedings
  of the 28th International Conference on Evaluation and Assessment in Software
  Engineering}, 2024, pp. 201--210.

\bibitem[Terrell et~al.(2017)Terrell, Kofink, Middleton, Rainear, Murphy-Hill,
  Parnin, and Stallings]{terrell2017gender}
\BIBentryALTinterwordspacing
J.~Terrell, A.~Kofink, J.~Middleton, C.~Rainear, E.~Murphy-Hill, C.~Parnin, and
  J.~Stallings, ``{Gender differences and bias in open source: Pull request
  acceptance of women versus men},'' \emph{PeerJ Computer Science}, vol.~3, p.
  e111, 2017. [Online]. Available:
  \url{https://doi.org/10.7287/peerj.preprints.1733v2}
\BIBentrySTDinterwordspacing

\bibitem[Kofink(2015)]{kofink2015contributions}
A.~Kofink, ``Contributions of the under-appreciated: gender bias in an
  open-source ecology,'' in \emph{Companion Proceedings of the 2015 ACM SIGPLAN
  International Conference on Systems, Programming, Languages and Applications:
  Software for Humanity}, 2015, pp. 83--84.

\bibitem[Canedo et~al.(2020)Canedo, Bonif{\'a}cio, Okimoto, Serebrenik, Pinto,
  and Monteiro]{canedo2020work}
E.~D. Canedo, R.~Bonif{\'a}cio, M.~V. Okimoto, A.~Serebrenik, G.~Pinto, and
  E.~Monteiro, ``Work practices and perceptions from women core developers in
  oss communities,'' in \emph{International Symposium on Empirical Software
  Engineering and Measurement}, 2020, pp. 1--11.

\bibitem[Ford et~al.(2019)Ford, Behroozi, Serebrenik, and
  Parnin]{ford2019beyond}
D.~Ford, M.~Behroozi, A.~Serebrenik, and C.~Parnin, ``Beyond the code itself:
  how programmers really look at pull requests,'' in \emph{ICSE-SEIS
  2019}.\hskip 1em plus 0.5em minus 0.4em\relax IEEE, 2019, pp. 51--60.

\bibitem[Singh(2019)]{singh2019women}
V.~Singh, ``Women-only spaces of open source,'' in \emph{2019 IEEE/ACM 2nd
  International Workshop on Gender Equality in Software Engineering
  (GE)}.\hskip 1em plus 0.5em minus 0.4em\relax IEEE, 2019, pp. 17--20.

\bibitem[Melaku et~al.(2020)Melaku, Beeman, Smith, and
  Johnson]{melaku2020better}
T.~M. Melaku, A.~Beeman, D.~G. Smith, and W.~B. Johnson, ``Be a better ally,''
  \emph{Harvard Business Review}, vol.~98, no.~6, pp. 135--139, 2020.

\bibitem[Alami et~al.(2024)Alami, Zahedi, and Ernst]{alami2024you}
A.~Alami, M.~Zahedi, and N.~Ernst, ``Are you a real software engineer? best
  practices in online recruitment for software engineering studies,'' in
  \emph{International Workshop on Methodological Issues with Empirical Studies
  in Software Engineering}, 2024.

\bibitem[Danilova et~al.(2021)Danilova, Naiakshina, Horstmann, and
  Smith]{danilova2021you}
A.~Danilova, A.~Naiakshina, S.~Horstmann, and M.~Smith, ``Do you really code?
  designing and evaluating screening questions for online surveys with
  programmers,'' in \emph{ICSE 2021}.\hskip 1em plus 0.5em minus 0.4em\relax
  IEEE, 2021, pp. 537--548.

\bibitem[Stol and Fitzgerald(2018)]{stol2018abc}
K.-J. Stol and B.~Fitzgerald, ``The abc of software engineering research,''
  \emph{ACM Transactions on Software Engineering and Methodology (TOSEM)},
  vol.~27, no.~3, p.~11, 2018.

\end{thebibliography}

\end{document}